# A LEAN METHANE PREMIXED LAMINAR FLAME

# DOPED WITH COMPONENTS OF DIESEL FUEL

# PART I: N-BUTYLBENZENE

E. POUSSE, P.A. GLAUDE, R. FOURNET, F.BATTIN-LECLERC[*]

Département de Chimie-Physique des Réactions,

Nancy Université, CNRS, ENSIC,

1 rue Grandville, BP 20451, 54001 NANCY Cedex, France

Full-length article

[*] E-mail : Frederique.Battin-Leclerc@ensic.inpl-nancy.fr ; Tel.: 33 3 83 17 51 25 , Fax : 33 3 83 37 81 20

To better understand the chemistry involved during the combustion of components of diesel fuel, the structure of a laminar lean premixed methane flame doped with n-butylbenzene has been investigated. The inlet gases contained 7.1% (molar) of methane, 36.8% of oxygen and 0.96% of n-butylbenzene corresponding to an equivalence ratio of 0.74 and a ratio $C_{10}H_{14}$ / $CH_4$ of 13.5%. The flame has been stabilized on a burner at a pressure of 6.7 kPa using argon as diluent, with a gas velocity at the burner of 49.2 cm/s at 333 K. Quantified species included the usual methane $C_0$-$C_2$ combustion products, but also 16 $C_3$-$C_5$ hydrocarbons, 7 $C_1$-$C_3$ oxygenated compounds, as well as 20 aromatic products, namely benzene, toluene, phenylacetylene, styrene, ethylbenzene, xylenes, allylbenzene, propylbenzene, cumene, methylstyrenes, butenylbenzenes, indene, indane, naphthalene, phenol, benzaldehyde, anisole, benzylalcohol, benzofuran, and isomers of $C_{10}H_{10}$ (1-methylindene, dihydronaphtalene, butadienylbenzene). A new mechanism for the oxidation of n-butylbenzene is proposed whose predictions are in satisfactory agreement with measured species profiles in flames and flow reactor experiments. The main reaction pathways of consumption of n-butylbenzene have been derived from flow rate analyses.

**INTRODUCTION**

If many detailed kinetic models are available for the oxidation of mixtures representative of gasolines, they are much less numerous in the case of diesel fuels because of their more complex composition. The constituents of diesel fuel contain from 10 to 20 carbon atoms and include about 30% (by mass) of alkanes, the remaining part being mainly alkylcyclohexanes (24%), alkyldecalines (15%), alkylbenzenes (10%) and polycyclic naphtenoaromatic compounds [1]. If the oxidation of alkanes has been extensively studied, the abundance of models diminishes considerably when other families of components are considered and very few models exist for substituted cycloalkanes and aromatics compounds, except for toluene [2]. The oxidation of alkylbenzenes with alkyl side-chains from $C_2$ to $C_4$ has been studied in a flow reactor in Princeton [3-6] and that of n-propylbenzene in a jet-stirred reactor at Orleans [7]. The autoignition of n-butylbenzene has also been investigated in a rapid compression machine in Lille [8-10]. While premixed flames of ethylbenzene [11] and non-premixed methane flames doped with ethylbenzene and isomers of propylbenzene and butylbenzene [12] have been studied, no measurement in a premixed flame containing butylbenzene has yet been reported.

The first purpose of the present paper is to experimentally investigate the structure of a premixed laminar methane flame doped with n-butylbenzene. The use of a methane flame will allow us to be more representative of the combustion mixtures containing large hydrocarbons, such as those present in a diesel fuel, than hydrogen or unsaturated $C_2$ flames. Large alkanes decompose to methyl radicals and the use of a methane flame will capture the involved chemistry. This study will be performed using a lean flame for the chemistry to be better representative of that occurring in engines controlled via Homogenous Charge Combustion Ignition (HCCI), which are under development. The second objective is to use these results in order to develop a new mechanism for the oxidation of n-butylbenzene based on our experience in modeling the reactions of both alkanes [13] and light aromatic compounds (benzene [14] and toluene [15]).

## EXPERIMENTAL PROCEDURE

The experiments were performed using an apparatus developed in our laboratory to study temperature and stable species profiles in a laminar premixed flat flame at low pressure and recently used in the case of rich methane flames doped by light unsaturated soot precursors [16-18]. A scheme of the apparatus is presented in figure 1.

FIGURE 1

The body of the flat flame matrix burner, provided by McKenna Products, was made of stainless steel, with an outer diameter of 120 mm and a height of 60 mm (without gas/water connectors). This burner was built with a bronze disk (95% copper, 5% tin). The porous plate (60 mm diameter) used to assist the flame stabilization was water cooled (water temperature: 333 K) with a cooling coil sintered into the plate. The burner could be operated with an annular co-flow of argon to favor the stabilization of the flame.

This horizontal burner was housed in a water-cooled vacuum chamber evacuated by two primary pumps and maintained at 6.7 kPa by a regulation valve. This chamber was equipped of four quartz windows for optical access, a pressure transducer (MKS 0-100 Torr), a microprobe for samples taking and a thermocouple for temperature measurements. The burner could be vertically translated, while the housing and its equipments were kept fixed. A sighting telescope measured the position of the burner relative to the probe or the thermocouple with an accuracy of 0.01 mm. The flame was lit using an electrical discharge.

Gas flow rates were regulated by RDM 280 Alphagaz and Bronkhorst (El-Flow) mass flow regulators. Methane (99.95 % pure) was supplied by Alphagaz - Air Liquide. Oxygen (99.5% pure) and argon (99.995% pure) were supplied by Messer. Liquid butylbenzene, supplied by Alfa Aesar (purity 99%), was contained in a glass vessel pressurized with argon. After each load of the vessel, argon bubbling and vacuum pumping were performed in order to remove oxygen traces dissolved in the liquid hydrocarbon fuel. The liquid reactant flow rate was controlled by using a liquid mass flow controller, mixed to the carrier gas and then evaporated by passing through a single pass heat exchanger, the temperature of which was set above the boiling point of the mixture. Carrier gas flow rate was controlled by a gas mass flow controller located before the mixing chamber.

Temperature profiles were obtained using a PtRh (6%)-PtRh (30%) type B thermocouple (diameter 200 $\mu$m). The thermocouple wire was supported by an arm and crossed the flame horizontally to avoid conduction heat losses. The junction was located at the centre of the burner. The thermocouple was coated with an inert layer of BeO-$Y_2O_3$ to prevent catalytic effects [19]. The ceramic layer was obtained by dipping the thermocouple in a hot solution of $Y_2(CO_3)_3$ (93% mass.) and BeO (7% mass.) followed by drying in a Meker burner flame. This process was repeated about ten times until the whole metal was covered. Radiative heat losses are corrected using the electric compensation method [20].

The sampling probe was constructed of quartz with a hole of about 50 $\mu$m diameter ($d_i$). The probe was finished by a small cone with an angle to the vertical of about 20°. For temperature measurements in the flames perturbed by the probe, the distance between the junction of the thermocouple and the end of the probe was taken equal to two times $d_i$, i.e. to about 100 $\mu$m. The sampling quartz probe was directly connected to a heated transfer line made with a passivated stainless steel tube and heated at 423 K. This line was itself connected to a heated pressure transducer (MKS 0-100 Torr) and through heated valves to a turbomolecular pump, a pyrex line and a heated 10 ml stainless steel loop located inside a gas chromatograph. The pressure in the transfer line was always below 1.3 kPa (10 Torr) so that the pressure drop between the flame and the inlet of the probe ensured rapid quenching of reactions. The on-line connection to a chromatograph was made via a heated transfer line in order to analyse compounds above $C_6$.

The analyses could be then performed according to two methods:

➢ Gas samples of compounds with a sufficient vapour pressure were directed through the pyrex line towards a volume which was previously evacuated by the turbo molecular pump down to $10^{-7}$ kPa and which was then filled up to a pressure of 1.3 kPa and collected in a pyrex loop after compression by a factor 5 by using a column with rising mercury. Pressures in the pyrex line were measured by a MKS 0-10 Torr pressure transducer before compression and by a MKS 0-1000 Torr pressure transducer after compression. A chromatograph with a Carbosphere packed column and helium or argon as carrier gas was used to analyse $O_2$, $H_2$, CO and $CO_2$ by thermal conductivity detection and $CH_4$, $C_2H_2$, $C_2H_4$, $C_2H_6$ by flame ionisation detection (FID). Water was detected by TCD but not quantitatively analyzed. Calibrations were performed by analysing a range of samples containing known pressures of each pure compound to quantify. Mole fractions were derived from the known total pressure of the gas included in the pyrex loop. This method of sampling could allow us to analyze heavier hydrocarbons from allene to toluene using a chromatograph with a Haysep packed column with FID and nitrogen as gas carrier gas as in our previous study [18], but it was found more accurate and faster to use the second method described hereafter. For the compounds for which both methods could be used, a very good agreement was obtained between both measurements.

➢ Gas samples were directed towards the loop located inside the on-line gas chromatograph and previously evacuated by the turbo molecular pump down to $10^{-7}$ kPa. The loop was then filled up to a pressure of 1.3 kPa and contents were immediately analysed using FID and helium as gas carrier gas by using a capillary column (a HP-Plot Q or a HP-1 column). This method of sampling was used to analyze methane, all the hydrocarbons from $C_3$ and the oxygenated products other than carbon oxides.

Figure 2 presents a typical chromatogram obtained with the HP-Plot Q column. The identification of these compounds was performed using a gas chromatograph with mass spectrometry detection (GC-MS) fitted with the same column or by comparison of retention times when injecting the product alone in gas phase. Four $C_3$ species (propyne, allene, propene and propane), nine $C_4$ species (diacetylene, vinylacetylene, 1-butyne, 2-butyne, 1,3-butadiene, 1,3-butadiene, 1-butene, iso-butene and butane) and four $C_5$ species (cyclopentene, cyclopentadiene, 1,3-pentadiene and isoprene (2-methyl-1,3-butadiene) were observed. The butenes, mainly 1-butene and iso-butene, were not well separated. In this lean flame, it was also possible to quantify seven oxygenated species (methanol, ketene, acetaldehyde, ethanol, acrolein, propanal and acetone). For methane, $C_3$ species, 1,3-butadiene and butenes, calibrations were performed by analyzing each pure compound for a range of known pressures in the transfer line and mole fractions were derived from the known total pressure. For other species, we have used the fact that the response of an FID detector is linked to the Effective Carbon Number (ECN) in the molecule [21]. The detector response coefficient ($DRC_{GP-i}$) of each compound $i$ can be calculated from the response of a molecule of close structure ($DRC_{GP-mod}$), the ECN of this molecule ($ECN_{mod}$) and that of the compound $i$ ($ECN_i$):

$$DRC_{GP-i} = DRC_{GP-mod} \times (ECN_i / ECN_{mod})$$

ECN is obtained as the sum of the contributions of all the atoms of carbon and oxygen included in the molecules. These contributions have been taken to 1 for aliphatic and aromatic atoms of carbon, 0.95 for olefinic ones, 1.3 for acetylenic ones and 0 for carbonyl ones (C=O) and to -1, -0.6 and -0.25 for atoms of oxygen included in ethers, in primary alcohols and in tertiary alcohols, respectively [21].

<center>FIGURE 2</center>

Figure 3 presents a typical chromatogram obtained with the HP-1 column. The identification of these compounds was performed using a gas chromatograph with GC-MS detection fitted with the same column. Two $C_6$ cyclic species (methycyclopentene and methylcyclopentadiene), twelve monocyclic aromatic compounds (benzene, toluene, xylenes, phenylacetylene, ethylbenzene, styrene, allylbenzene, n-propylbenzene, cumene (iso-propyl-benzene), methylstyrenes, butenylbenzenes (compounds with $C_{10}H_{12}$ as chemical formula) and butadienylbenzene) and 5 bicyclic aromatic compounds (indene, indane, methylindene, dihydronaphthalene and naphthalene) were identified as the most probable products according to their mass spectrum. Several peaks have been observed for xylenes, methylstyrenes and compounds with $C_{10}H_{10}$ as chemical formula, which have been identified as butadienylbenzenes, methylindene and dihydronaphthalene. In this last case, the separate quantification of the different isomers was not possible. Five oxygenated aromatic compounds (phenol, benzaldehyde, benzylalcohol, anisole (methylphenylether) and benzofuran) were also quantified. For those compounds having a low vapour pressure, calibrations were performed in two steps. First, a calibration was made for gas-phase cyclohexane, a cyclic compound with a much larger vapour pressure, by analysing this species for a range of known pressures introduced in the transfer line. This calibration allowed the detector response coefficient of gas-phase cyclohexane ($DRC_{GP\text{-}CHX}$) to be obtained:

$$DRC_{GP\text{-}CHX} = A_{GP\text{-}CHX} / P_{GP\text{-}CHX}$$

with:

$A_{GP\text{-}CHX}$: the area of the cyclohexane peak in the case of the gas-phase injection,

$P_{GP\text{-}CHX}$: the cyclohexane partial pressure in the transfer line.

The detector response coefficient ($DRC_{GP-P}$) of each gas-phase aromatic compound of interest, P, was derived from its relative detector response coefficient ($DRC_{L-P}$) compared to that of cyclohexane ($DRC_{L-CHX}$) obtained by injecting an acetone solution containing known mass of both cyclohexane ($m_{CHX}$) and the P ($m_P$) in liquid phase:

$$DRC_{GP-P} = (DRC_{L-P} / DRC_{L-CHX}) \times DRC_{GP-CHX} \times (M_P / M_{CHX})$$

with:

$DRC_{L-CHX} = A_{L-CHX} / m_{CHX}$, $A_{L-CHX}$ is the area of the cyclohexane peak in the case of the liquid injection,

$DRC_{L-P} = A_{L-P} / m_P$, $A_{L-P}$ is the area of the P peak in the case of the liquid injection,

$M_{CHX}$ the molar mass of cyclohexane,

$M_P$ the molar mass of P.

This method of calibration has been used for all major aromatic compounds, except from methylstyrenes, butenylbenzenes, compounds with $C_{10}H_{10}$ as chemical formula, phenol, benzylalcohol and benzofuran for which the ECN method has been used.

FIGURE 3

Calculated uncertainties on the species quantifications were about ± 5% for major hydrocarbon compounds and ± 10% for oxygen, hydrogen, carbon oxides and the minor hydrocarbon products. The detection limit of the FID is about 2 ppm.

**EXPERIMENTAL RESULTS**

A laminar premixed flat flame has been stabilized on the burner at 6.7 kPa (50 Torr) with a gas flow rate of 5.44 l/min corresponding to a gas velocity at the burner of 49.2 cm/s at 333 K and with mixtures containing 7.1% (molar) of methane, 36.8% of oxygen and 0.96% of n-butylbenzene corresponding to an equivalence ratio of 0.74.

Figure 4 displays the experimental temperature profiles obtained with and without the probe showing that the presence of the probe induces a thermal perturbation involving a lower temperature. Without the probe, the lowest temperatures measured the closest to the burner (0.47 mm above) were around 1095 K. Due to the thinness of this lean flame and the size of the thermocouple, it was not possible to measure lower temperatures. The highest temperatures were reached from 5 mm above the burner and were around 1960 K.

FIGURE 4

Figure 5 presents the profiles of both hydrocarbon reactants and shows that n-butylbenzene (fig 5b) is consumed close to the burner, at 1 mm height, while some methane (Fig. 5a) remains up to 2 mm.

FIGURE 5

Figures 6 and 7 present the profiles of oxygen (fig 6a), hydrogen (fig 7a), water (fig. 6b) and of the main $C_0$-$C_2$ species involved in the combustion of methane versus the height above the burner. The mole fraction of water has been obtained from a material balance on the other major species. In this lean flame, the profiles of carbon monoxide (fig. 6c) and hydrogen display a marked maximum at 2 mm height and the major final products are for a large extent carbon dioxide (fig. 6d) and water. Ethylene (fig. 7c) is the most abundant $C_2$ species and is produced first. It reaches its maximum concentration close to the burner around 1 mm. The profiles of ethane (fig. 7d) and of acetylene (fig. 7b) peak around 1.1 mm and 1.2 mm, respectively.

FIGURES 6 AND 7

Figure 8 presents the profiles of the observed $C_3$ products, with propene (fig. 8c) and propane (fig. 8d) peaking first around 1 mm above the burner, while the maxima for allene (fig. 8b) and propyne (fig. 8a) are around 1.25 mm. The peak mole fractions observed for propene and propane are more than ten times larger than that of allene and propyne.

## FIGURE 8

Figure 8 displays the profiles of $C_4$ species. 1-butyne (fig. 9c), 1-2 butadiene (fig. 9f), butenes (fig. 9g) and butane (fig. 9h) are produced first and reach their maximum concentration close to the burner, around 1.1 mm. The profiles of diacetylene (fig. 9a), vinylacetylene (fig. 9b), 2-butyne (fig. 9d) and 1,3-butadiene (fig. 9e) peak around 1.5 mm. The most abundant $C_4$ compounds are vinylacetylene, 1,3-butadiene and butanes with peak concentrations between 60 and 80 ppm. Butynes and 1,2-butadiene are present in very low amounts with peak concentrations well below 10 ppm.

## FIGURE 9

Figure 10 presents the profiles of $C_5$-$C_6$ non-aromatic species. Cyclopentene (fig. 10b), 1,3-pentadiene (fig. 10c) and isoprene (fig. 10d) reach their maxima first around 1 mm above the burner. The profiles of methylcyclopentadiene (fig. 10e) peak further around 1.1 and the maxima of that of cyclopentadiene (fig. 10a), the most abundant species of this series with a peak concentration of 50 ppm, and methylclopentene (fig. 10f) are the last ones around 1.5 mm.

## FIGURE 10

Figure 11 displays the profiles of light oxygenated species. Methanol (fig. 11a), ketene (fig. 11b), acetaldehyde (fig. 11c) and ethanol (fig. 11d), which are intermediate products of the combustion of methane, as well as propanal (fig. 11f), are produced very early and reach their maximum concentration close to the burner, around 0.9 mm. The profiles of acroleïn (fig. 11e) and acetone (fig. 11g) peak around 1.2 mm. The most abundant of these species are methanol and acetaldehyde with peak concentrations above 150 ppm.

## FIGURE 11

Figures 12 and 13 present the profiles of monocyclic aromatic products. The maxima of the profiles of toluene (fig. 12b), xylenes (fig. 12c), ethylbenzene (fig. 12f), styrene (fig. 12e) allylbenzene (fig. 13a), n-propylbenzene (fig. 13b) and butenylbenzenes (fig. 13e) occur first around 1 mm above the burner, followed by that of benzene (fig. 12a), cumene (fig. 13c) and methylstyrenes (fig. 13d) around 1.3 mm and finally that of phenylacetylene (fig. 12d) around 1.5 mm. The most abundant of these species are benzene with a peak concentration of 800 ppm, toluene with a peak concentration of 800 ppm and styrene with a peak concentration of 1000 ppm)and to a lesser extent allylbenzene and butenylbenzenes with peaks concentration  about 100 ppm. Xylenes, phenylacetylene and  n-propylbenzene are present in low amounts with peak concentrations below 15 ppm.

FIGURES 12 AND 13

Figure 14 displays the profiles of bicyclic aromatic compounds. The maxima of the profiles of indene (fig. 14a) and the isomers of $C_{10}H_{10}$ (fig. 14d) are observed first around 1 mm above the burner, followed by that of indane (fig. 14b) around 1.3 mm and finally that of naphthalene (fig. 14c) around 1.8mm. The most abundant of these species is indene with a peak concentration of 50 ppm. Indane and naphthalene are present in very low amounts with peak concentrations well below 10 ppm.

FIGURE 14

Finally figure 15 presents the profiles of oxygenated aromatic species. Benzaldehyde (fig. 15b), the most abundant of these species with a peak concentration of 150 ppm, and benzylalcohol (fig. 15c) reach their maxima first around 1 mm above the burner. The profiles of phenol (fig. 15a) and anisole (fig. 15d) peak further around 1.4 mm and the maximum of that of benzofuran (fig. 15e) is the last one around 1.5 mm. This last compound, which can be a precursor of dioxins, corresponds to a very low peak concentration of 2 ppm.

**DESCRIPTION OF THE PROPOSED MECHANISM**

The mechanism proposed here to model the oxidation of n-butylbenzene includes the previous mechanisms that were built by our team to model the oxidation of $C_3$-$C_5$ unsaturated hydrocarbons [22, 23, 16-18] benzene [14] and toluene [15]. Thermochemical data were estimated by the software THERGAS developed in our laboratory [24], which is based on the group additivity methods proposed by Benson [25], apart from the heat of formation of biaromatic species which were taken from Burcat and Ruscic [26]. The complete mechanism is available as supplementary material.

*Reaction base for the oxidation of $C_3$-$C_5$ unsaturated hydrocarbons [22, 23, 16-18]*

This $C_3$-$C_5$ reactions database was built from a review of the literature and is an extension of our previous $C_0$-$C_2$ reactions database [27]. This $C_0$-$C_2$ reactions database includes all the unimolecular or bimolecular reactions involving radicals or molecules including carbon, hydrogen and oxygen atoms and containing less than three carbon atoms. The kinetic data used in this base were taken from the literature and are mainly those proposed by Baulch *et al.* [28] and Tsang *et al.* [29].

The $C_3$-$C_5$ reactions database [22-23] includes reactions involving $C_3H_2$, $C_3H_3$, $C_3H_4$ (allene and propyne), $C_3H_5$, $C_3H_6$, $C_4H_2$, $C_4H_3$, $C_4H_4$, $C_4H_5$, $C_4H_6$ (1,3-butadiene, 1,2-butadiene, 1-butyne and 2-butyne), $C_4H_7$ (6 isomers), some linear and branched $C_5$ compounds and well as cyclopentene and derived species, and the formation of benzene and toluene.

In this reactions base, pressure-dependent rate constants follow the formalism proposed by Troe [30] and efficiency coefficients have been included. It has been validated by modeling experimental results obtained in a jet-stirred reactor for methane and ethane, profiles in laminar flames of methane, acetylene and 1,3-butadiene and shock tube autoignition delay times for acetylene, propyne, allene, 1,3-butadiene, 1-butyne and 2-butyne. The agreement in modeling concentration profiles in laminar flames of pure methane is satisfactory at an equivalence ratio of 1.55 [16], but deteriorates for the formation of ethylene and acetylene at an equivalence ratio of 0.4. An improved version has recently been used to model the structure of a laminar premixed flame of methane doped with allene, propyne, 1,3-butadiene and cyclopentene [16-18]. Compared to this last version, reactions of formation and consumption of acetone have been added.

*Mechanisms for the oxidation of benzene [14] and toluene [15]*

Our mechanism for the oxidation of benzene contains 135 reactions and includes the reactions of benzene and of cyclohexadienyl, phenyl, phenylperoxy, phenoxy, hydroxyphenoxy, cyclopentadienyl, cyclopentadienoxy and hydroxycyclopentadienyl free radicals, as well as the reactions of ortho-benzoquinone, phenol, cyclopentadiene, cyclopentadienone and vinylketene, which are the primary products yielded. Validations have been made using for comparison experimental results obtained in a jet-stirred and a plug flow reactors, profiles in a laminar flame of benzene and shock tube autoignition delay times.

The mechanism for the oxidation of toluene contains 193 reactions and includes the reactions of toluene and of benzyl, tolyl, peroxybenzyl (methylphenyl), alcoxybenzyl and cresoxy free radicals, as well as the reactions of benzaldehyde, benzyl hydroperoxyde, cresol, benzylalcohol, ethylbenzene, styrene and bibenzyl. Validations have been made using for comparison experimental results obtained in a jet-stirred and a plug flow reactors, and shock tube autoignition delay times.

Compared to these published mechanisms, reactions of formation and consumption of some products have been added. Cumene is obtained by combination of methyl and phenylethyl radicals or of phenyl and iso-propyl radicals. Methylstyrene is produced by addition of methyl radicals to styrene or by combination of phenylvinyl and methyl radicals. Anisole and benzofuran both derive from phenoxy radicals.

*Mechanism proposed for the oxidation of n-butylbenzene*

Table I presents the reactions of n-butylbenzene and derived species. The complete mechanism is then composed of this submechanism and of the submechanisms described above. It involves 210 species in 1478 reactions and can be used to run simulations using CHEMKIN softwares [31]. Table II presents the names, the formulae and the heats of formation of the aromatic species included in this mechanism and containing at least 9 atoms of carbon.

<div align="center">TABLES I AND II</div>

The primary mechanism contains the reactions of n-butylbenzene and of the radicals directly deriving from it. We have considered the unimolecular reactions of n-butylbenzene, the ipso-additions and the H-abstractions by oxygen molecules and radicals present with important concentrations. The rate constants of the unimolecular decompositions involving the breaking of a C-C bond (reactions 1-4) have been calculated from the modified collision theory and thermokinetic relationships [25]. The rate constants of the unimolecular decompositions to give H atoms and phenylbutyl radicals (reactions 5-8) have been deduced from that of the reverse reaction, $k = 1.0 \times 10^{14}$ s$^{-1}$ according to Allara *et al.* [34]. The kinetic parameters of the bimolecular initiations with oxygen molecules (reactions 9-12) have been obtained using the correlation proposed by Ingham *et al.* [35]. The rate constants of the ipso-additions of hydrogen and oxygen atoms and of hydroxyl and methyl radicals (reactions 13-17) have been deduced from value proposed for the similar reactions in the cases of benzene and toluene. The rate constants for the abstractions of alkylic H-atoms (reactions 19-21, 23-25, 27-29, 31-33, 35-37) were deduced from the correlations proposed by Buda *et al.* for alkanes [13] and those for the abstractions of allylic H-atoms (18, 22, 26, 30, 34, 38-41) were deduced from the correlations proposed by Heyberger *et al.* [39] and by analogy with similar reactions of ethylbenzene [15].

The reactions of phenybutyl and phenylpropyl radicals were mainly derived from the reactions of alkyl and alkenyl radicals generated by EXGAS software [13, 33, 41-44]. In the case of phenylbutyl radicals, reactions involved isomerizations (reactions 42-46), decompositions by breaking of a C-C bond to form styrene, ethylene, propene and 1-butene as stable molecules, the formation of phenylbutenes by breaking a C-H bond (reactions 48, 51-52, 54-55 and 57) or by oxidation with oxygen molecules (reactions 58 to 63). Termination steps (reactions 64 to 67) were only written for the resonance stabilized 4-phenylbut-1-yl radicals: combinations with $HO_2$ radicals led to phenylbutoxy radicals, combinations with $CH_3$ radicals to 2-phenyl-n-pentane and disproportionnations with benzyl and allyl radicals to 1-phenyl-1-butene, toluene and propene, respectively. Phenylpropyl radicals were considered to react by isomerisation followed by decomposition (reaction 68) to lead ultimately to styrene, by cyclisation (reaction 69) to give indane, by β-scission decompositions (reactions 70-71) yielding ethylene and allylbenzene as stable molecules, by oxidation (reaction 72) to produce allylbenzene and by combination with H-atoms to form propylbenzene (reaction 73). The reactions of 1-butyl radicals (reactions 74 to 78) are a reduced version of those generated by EXGAS software for their high temperature oxidation.

The reactions of oxygenated aromatic radicals are still very uncertain. The globalized reactions considered for butylphenoxy radicals (reactions 79-80) are derived from those of phenoxy and started as an elimination of carbon monoxide. Propylbenzylalcoxy radicals have been assumed to react by isomerizations with a rapid decomposition of the obtained radicals (reactions 81-82) or by direct β-scission decompositions (reactions 83-84).

The secondary mechanism includes the reactions of the primary products which are not considered in the mechanisms of the oxidation of benzene and toluene, namely the three isomers of phenylbutene (reactions 85 to 164), allylbenzene (reactions 165 to 182), propylbenzene (reactions 183 to 200), 2-phenyl-n-pentane (reactions 201 to 204) and indane (reactions 205 to 216), as well as those of the derived products, namely butadienylbenzenes (reactions 217 to 226), the bicyclic isomers of $C_{10}H_{10}$ (reactions 227 to 231), indene (reactions 232 to 243) and naphthalene (reactions 244 to 255).

In the case of the isomers of phenylbutene and allylbenzene, we have considered the bimolecular initiations with oxygen molecules, the ipso-additions of H-atoms and methyl radicals to give benzene and toluene, the additions to the double bond followed by the decompositions of the obtained adducts and the H-abstractions to give phenylalkenyl radicals, which can react by decompositions by β-scission, oxidations, cyclizations and combinations. The rate constants of the additions to the double bond and of the formation and the consumption of alkenyl radicals were derived from our previous work on the oxidation of alkenes [39, 41, 44]. The cyclizations of phenylbutenyl and phenylpropenyl radicals lead to bicyclic isomers of $C_{10}H_{10}$ (dihydronaphthalene and methylindene) and indene, respectively. Their rate constant has been assumed equal to that proposed by Gierzak [46] for the cyclization of pentenyl radicals to give cyclopentyl radicals. For propylbenzene, unimolecular and bimolecular initiations, ipso-additions of H-atoms and methyl radicals and H-abstractions have been written. In the case of 2-phenyl-pentane, which has not been experimentally observed, only four globalized H-abstractions were taken into account. Indane has been considered as yielding indene by bimolecular initiations with oxygen molecules and H-abstractions by small radicals followed by β-scission decompositions of the obtained indanyl radicals.

As only very few studies concerning their oxidation have been published, the reactions of the other products are still very uncertain. For butadienylbenzene, globalized reactions have been written starting by bimolecular initiations with oxygen molecules or H-abstractions by O- and H-atoms and OH radicals, both followed by the cyclization of the obtained radicals yielding naphthalene. The reactions of the bicyclic isomers of $C_{10}H_{10}$ have also been assumed as leading to naphthalene by molecular dehydrogenation or by H-abstractions. In the case of indene, the addition of OH radicals to the double bond included in the five members cycle has been written, as well as bimolecular initiations with oxygen molecules and H-abstractions by small radicals to give resonance stabilized indenyl radicals, the combinations of which has been assumed to be similar to those written by Da Costa *et al.* [14] for cyclopentadienyl radicals. Finally, as previously assumed in the work of Bounaceur *et al.* [51], the reactions of naphthalene and naphthyl radicals can be derived from those of benzene and phenyl radicals.

**COMPARISON BETWEEN EXPERIMENTAL AND SIMULATED RESULTS**

Simulations were performed using PREMIX from CHEMKIN II [31] using the experimental temperature profile as an input. To compensate the perturbations induced by the quartz probe and the thermocouple, the temperature profile used in calculations is an average between the experimental profiles measured with and without the quartz probe, shifted 0.4 mm away from the burner surface, as shown in figure 4.

Figures 5 to 7 and 11 show that the model reproduces satisfactorily the consumption of reactants and the formation of the main $C_0$-$C_2$ products, including the oxygenated ones, related to the consumption of methane in the flame doped with n-butylbenzene. Only the formation of hydrogen in the burned gas is largely overestimated, that of ethylene is underpredicted by a factor almost 2 and that of ketene and ethanol are overpredicted by factors 10 and 20, respectively.

To decouple the effect due to the increase of equivalence ratio ($\phi$) and that induced by the presence of n-butylbenzene, figures 5 to 7 and 11 display also the results of a simulation performed for a flame containing 7.1% methane and 15.64 % oxygen (with no additive) for $\phi=0.74$, i.e. equal to that of the doped flame. As the temperature rise is mainly influenced by $\phi$, we have used the same temperature profile as that used to model the doped flame. The profile of methane is very similar for the doped and undoped flames at the same $\phi$. However the content in atoms of carbon is 2.35 lower in the pure methane flame compared to the doped one (the C/O ratio is equal to 0.096 in the undoped flame and to 0.225 in the seeded one) which is well reflected by the profiles of carbon oxides and ethane. While the profile of ethane is not much affected, the formation of ethylene and acetylene is increased by the addition of n-butylbenzene by a factor much larger than the increase in the number of atoms of carbon showing clearly to what extent these products derive from the decomposition of this additive. The presence of the additive does not influence the formation of methanol, but increase also considerably that of ketene, acetaldehyde and ethanol.

Figures 8 to 11 present the comparison between experimental and simulated data for the other non-aromatic products, which are all modeled with a factor better than 2, except from n-butane and methylcyclopentene, for which a deviation of a factor 3 is observed, and isoprene, methylcyclopentadiene, and acetone, the formation of which is more considerably underestimated. Comparison with the simulation of a pure methane flame at $\Phi=0.74$ shows that $C_3$ compounds, which are formed in significant amounts in the doped flame, are almost non-existent in the lean mixture without additive, except for propane. Specific reactions leading to unsaturated $C_3$ products are then also induced by the presence of the cyclic additive.

Figures 12 to 15 display the comparison for the profiles of aromatic compounds. Concerning the most abundant species, the formation of benzene, toluene and allylbenzene is very well reproduced, that of styrene is underestimated by a factor 1.5, that of benzaldehyde is overestimated by the same factor and that of butenylbenzenes is overestimated by a factor 2.5.

The poor separation of the $C_{10}H_{12}$ isomers on the GC column that we used is a possible explanation for this discrepancy. While the simulated profiles of bicyclic compounds notably differ from the experimental ones, the maximum mole fractions obtained for minor products are all modeled with a factor better than 4, except from benzofuran, the formation of which is underestimated by a factor 10 showing that that a production pathway is missing in the model. The production of phenol is overestimated by a still larger factor.

It is worth noting that the largest disagreements are obtained for several oxygenated compounds. For ethanol, ketene and phenol which are strongly overestimated by the model, it is probable that, despite the heated transfer line, some of these oxygenated compounds may be absorbed on the walls and result in lower measured values than those predicted by our model. While the reactions of ketene are still very uncertain, those of ethanol and phenol are rather well determined with pressure effects taken into account.

In order to extend the temperature ranges of these simulations, the results of Litzinger *et al.* [6], which were obtained in a flow reactor at 1060 K, at atmospheric pressure, with nitrogen as bath gas, for an initial concentration of n-butylbenzene of 620 ppm and for an equivalence ratio of 0.98, have also been modeled. Figure 16 displays comparisons between the experimental and computed mole fraction of reactants and main products. This figure shows that a globally correct agreement can be observed. The consumptions of hydrocarbon and oxygen (fig. 16a) are correctly reproduced, as well as the formation of the major products, methane (fig. 16b), ethylene (fig. 16b), benzene (fig. 16c), toluene (fig. 16c), styrene (fig. 16c), allylbenzene (fig. 16c). The production of carbon monoxide (fig. 16b) is overestimated by a factor 1.5 at the longest residence times.

FIGURE 16

**DISCUSSION**

Figure 17 displays the main flows of consumption of the n-butylbenzene at a temperature about 1140 K corresponding to 87% reactant conversion. A large enough conversion has been chosen, so that the major ways of consumption of the primary products can be observed. Under these conditions, butylbenzene is mainly consumed by H-abstractions (40 % if its consumption) to give the resonance stabilized 4-phenylbut-4-yl radical which is the main source of styrene through a β-scission decomposition. At this temperature, this radical is consumed faster than it is produced which is only possible because of its accumulation at lower temperature.

FIGURE 17

The other important channels of consumption of n-butylbenzene are the H-abstractions leading to the other 4-phenylbutyl radicals (globally 41 % of its consumption). A part of 4-phenylbut-1-yl and 4-phenylbut-2-yl radicals isomerizes to give resonance stabilized 4-phenylbut-4-yl radicals. However 4-phenylbut-1-yl radicals mainly decompose to give ethylene and 2-phenyleth-1-yl radicals yielding phenyl radicals and ethylene or, for a smaller part, styrene and H-atoms. The decomposition of 4-phenylbut-1-yl radicals explains most part of the formation of ethylene. 4-phenylbut-2-yl radicals are consumed to produced allylbenzene and methyl radicals, butenylbenzene and H-atoms and 1-butene and phenyl radicals. This last channel is the main way to form 1-butene. Phenyl radicals are an important source of toluene and phenoxy radicals yielding phenol, and to a much lesser extent of benzene. Anisol and benzofuran are produced from minor reactions of phenoxy with methyl radicals and acetylene, respectively. As benzofuran is strongly underestimated by the model, it is probable that important way of formation of this compound are missing. Finally 4-phenylbut-3-yl radicals are almost completely consumed to form propene and benzyl radicals, which are the major source of toluene, benzaldehyde, ethylbenzene, benzylalcohol and n-propylbenzene. The formation of xylenes derive from toluene through the ipso-addition of methyl radicals. The largest part of the formation of propene is due to reactions of 4-phenylbut-3-yl radicals.

There are two additional minor channels of consumption of n-butylbenzene by ipso-additions. The ipso-addition of H-atoms (6 % of its consumption) is the main source of benzene. Butyl radicals are also produced through this way and mostly decompose to form ethylene and ethyl radicals and to a lesser extent 1-butene and H-atoms. A very small fraction of them also combines with H-atoms to give butane. Ethyl radicals are a source of acetaldehyde by reaction with O-atoms and of ethanol by combinations with OH radicals. The ipso-addition of O-atoms (3.7 % of its consumption) leads ultimately to carbon monoxide, as well as to cyclopentadienyl radicals and 1-butene or to H-atom, propene and benzene.

Styrene reacts mainly by additions of OH radicals or of H-atoms and by H-abstractions. The additions of OH radicals are a source of formaldehyde and benzyl radicals and of benzaldehyde and methyl radicals. The additions of H-atoms produce resonance stabilized 2-phenyleth-2-yl radicals, the combinations of which yield cumene and ethylbenzene. Phenylvinyl radicals which are obtained by H-abstractions from styrene react mainly with oxygen molecules yielding resonance stabilized benzoyl radicals which decompose to produce phenyl radicals and carbon monoxide. Minor channels consuming phenylvinyl radicals involve the formation of methylstyrene and of phenylacetylene.

Butenylbenzenes and allylbenzene react mainly by H-abstractions to give resonance stabilized radicals the cyclization of which leads to methylindene and indene, respectively. The cyclization of the alkylic alkenyl radicals, which are formed by another minor channel consuming butenylbenzenes, produces dihydronaphthalene. The cyclization of the phenylpropyl radicals, which are obtained by addition of H-atoms to allylbenzene, explains the formation of indane (not shown in fig. 17, as it corresponds to a very small flow rate). The flow rate of the addition to allylbenzene is much smaller than that of the H-abstractions explaining the lower production of indane compared to indene. The reactions of methylindene and dihydronaphthalene lead to naphthalene or to resonance stabilized cyclopentadienyl radicals. These cyclic $C_5$ radicals mostly yield cyclopenpentadiene by combination with H-atoms, but also, to a lower extent, methylcyclopentadiene by combinations with methyl radicals, and acetylene or 1,3-butadiene by opening of the ring. A small part of cyclopentadienyl radicals derives also from the CO elimination from phenoxy radicals. Cyclopentene and methylcyclopentene derive from cyclopentadiene and methylcyclopentadiene, respectively, by additions of H-atoms followed by combinations with H-atoms

Let us now describe the major ways of formation of the minor species which do not derives directly from butylbenzene. Ethane, propane and methanol are formed by combination of methyl radicals with themselves, ethyl and OH radicals, which explains why their formation is not much affected by the presence of butylbenzene. Allene derives from propene via the formation of resonance stabilized allyl radicals. Propyne is formed by addition of O-atoms to vinylacetylene, which is obtained from cyclopentadienone which derives from benzoquinone, a minor product of the reaction of phenyl radicals with oxygen molecules. The additions of methyl radicals to propyne yield iso-butyl radicals, the recombination of which is the main source of iso-butene. Diacetylene and butynes derive from vinylacetylene. 1,2-butadiene is obtained from resonance stabilized 1-buten-3-yl radicals, which are produced by H-abstractions from 1-butene. 1,3-pentadiene derives from cyclopentene after H-abstractions and opening of the ring and isoprene from the addition of vinyl radicals to allene. Ketene and acetone both derive from $CH_3CO$ radicals obtained by H-abstraction from acetaldehyde. Propanal is produced by addition of OH radicals to 1-butene or butenylbenzenes. Acroleïn is mainly obtained by reaction of allyl radicals with O-atoms.

**CONCLUSION**

This paper presents new experimental results for a lean premixed laminar flame of methane seeded with n-butylbenzene, as well as a new mechanism developed to reproduce the combustion of this substituted aromatic compound, which can be considered as a model molecule of an important class of components of diesel fuels. Profiles of temperature have been measured and mole fraction profiles have been obtained for 55 identified stable species from $C_0$ to $C_{10}$, including 20 aromatic products and 12 oxygenated compounds other than the reactants. Several of these species are considered as toxic pollutants, this is the case of oxygenated compounds (acroleïn). Several of these products are also known as soot precursors, this is the case of all the aromatic compounds.

Satisfactory agreement has been obtained between experimental results and simulations, apart from 4 oxygenated species (ethanol, ketene, phenol, benzofuran). The prediction of the profiles of bicyclic aromatic compounds (indene, indane, naphthalene, benzofuran) could also be improved when a better knowledge of their chemistry will be available.

**AKNOWLEDGEMENT**

This work has been supported by PSA Peugeot Citroën and TOTAL

**TABLE I: REACTIONS OF BUTYLBENZENE AND OF DERIVED SPECIES**

The rate constants are given (k=A $T^n$ exp(-$E_a$/RT)) in cc, mol, s, kcal units.

| Reactions | A | n | $E_a$ | Footnote | No |
|---|---|---|---|---|---|
| **Reactions of butylbenzene ($\Phi$-C$_4$H$_9$)** | | | | | |
| *Unimolecular initiations* | | | | | |
| $\Phi$-C$_4$H$_9$=C$_6$H$_5$#+C$_4$H$_9$ | $1.0\times10^{16}$ | 0.0 | 97.0 | a | (1) |
| $\Phi$-C$_4$H$_9$=benzyl+nC$_3$H$_7$ | $2.5\times10^{16}$ | 0.0 | 74.7 | b | (2) |
| $\Phi$-C$_4$H$_9$=$\Phi$-C$_2$H$_4$-1+C$_2$H$_5$ | $3.2\times10^{15}$ | 0.0 | 84.1 | b | (3) |
| $\Phi$-C$_4$H$_9$=$\Phi$-C$_3$H$_6$-1+CH$_3$ | $1.7\times10^{16}$ | 0.0 | 86.6 | b | (4) |
| *Decompositions by breaking of a C-H bond* | | | | | |
| $\Phi$-C$_4$H$_8$X+H=$\Phi$-C$_4$H$_9$ | $1.0\times10^{14}$ | 0.0 | 0.0 | c | (5) |
| $\Phi$-C$_4$H$_8$-2+H=$\Phi$-C$_4$H$_9$ | $1.0\times10^{14}$ | 0.0 | 0.0 | c | (6) |
| $\Phi$-C$_4$H$_8$-3+H=$\Phi$-C$_4$H$_9$ | $1.0\times10^{14}$ | 0.0 | 0.0 | c | (7) |
| $\Phi$-C$_4$H$_8$-4+H=$\Phi$-C$_4$H$_9$ | $1.0\times10^{14}$ | 0.0 | 0.0 | c | (8) |
| *Bimolecular initiations* | | | | | |
| $\Phi$-C$_4$H$_9$+O$_2$=$\Phi$-C$_4$H$_8$X+HO$_2$ | $1.4\times10^{12}$ | 0.0 | 35.1 | d | (9) |
| $\Phi$-C$_4$H$_9$+O$_2$=$\Phi$-C$_4$H$_8$-3+HO$_2$ | $1.4\times10^{13}$ | 0.0 | 49.9 | d | (10) |
| $\Phi$-C$_4$H$_9$+O$_2$=$\Phi$-C$_4$H$_8$-2+HO$_2$ | $1.4\times10^{13}$ | 0.0 | 49.2 | d | (11) |
| $\Phi$-C$_4$H$_9$+O$_2$=$\Phi$-C$_4$H$_8$-1+HO$_2$ | $2.1\times10^{13}$ | 0.0 | 52.3 | d | (12) |
| *Ipso-additions* | | | | | |
| $\Phi$-C$_4$H$_9$+H=benzene+C$_4$H$_9$ | $5.8\times10^{13}$ | 0.0 | 8.1 | e | (13) |
| $\Phi$-C$_4$H$_9$+CH$_3$=toluene+C$_4$H$_9$ | $1.2\times10^{12}$ | 0.0 | 15.9 | f | (15) |
| $\Phi$-C$_4$H$_9$+OH=phenol+C$_4$H$_9$ | $8.2\times10^{2}$ | 2.9 | 3.2 | g | (16) |
| $\Phi$-C$_4$H$_9$+H=OC$_6$H$_4$#C$_4$H$_9$+H | $1.7\times10^{13}$ | 0.0 | 3.6 | h | (17) |
| *Metatheses* | | | | | |
| $\Phi$-C$_4$H$_9$+O=$\Phi$-C$_4$H$_8$X+OH | $8.8\times10^{10}$ | 0.7 | 3.3 | i | (18) |
| $\Phi$-C$_4$H$_9$+O=$\Phi$-C$_4$H$_8$-3+OH | $2.6\times10^{13}$ | 0.0 | 5.2 | j | (19) |
| $\Phi$-C$_4$H$_9$+O=$\Phi$-C$_4$H$_8$-2+OH | $2.6\times10^{13}$ | 0.0 | 5.2 | j | (20) |
| $\Phi$-C$_4$H$_9$+O=$\Phi$-C$_4$H$_8$-1+OH | $5.1\times10^{13}$ | 0.0 | 7.9 | j | (21) |
| $\Phi$-C$_4$H$_9$+H=$\Phi$-C$_4$H$_8$X+H$_2$ | $5.4\times10^{4}$ | 2.5 | -1.9 | i | (22) |
| $\Phi$-C$_4$H$_9$+H=$\Phi$-C$_4$H$_8$-3+H$_2$ | $9.0\times10^{6}$ | 2.0 | 5.0 | j | (23) |
| $\Phi$-C$_4$H$_9$+H=$\Phi$-C$_4$H$_8$-2+H$_2$ | $9.0\times10^{6}$ | 2.0 | 5.0 | j | (24) |
| $\Phi$-C$_4$H$_9$+H=$\Phi$-C$_4$H$_8$-1+H$_2$ | $2.8\times10^{7}$ | 2.0 | 7.7 | j | (25) |
| $\Phi$-C$_4$H$_9$+OH=$\Phi$-C$_4$H$_8$X+H$_2$O | $3.0\times10^{6}$ | 2.0 | -1.5 | i | (26) |
| $\Phi$-C$_4$H$_9$+OH=$\Phi$-C$_4$H$_8$-3+H$_2$O | $2.6\times10^{6}$ | 2.0 | -0.8 | j | (27) |
| $\Phi$-C$_4$H$_9$+OH=$\Phi$-C$_4$H$_8$-2+H$_2$O | $2.6\times10^{6}$ | 2.0 | -0.8 | j | (28) |
| $\Phi$-C$_4$H$_9$+OH=$\Phi$-C$_4$H$_8$-1+H$_2$O | $2.6\times10^{6}$ | 2.0 | 0.5 | j | (29) |
| $\Phi$-C$_4$H$_9$+HO$_2$=$\Phi$-C$_4$H$_8$X+H$_2$O$_2$ | $6.4\times10^{3}$ | 2.6 | 12.4 | i | (30) |
| $\Phi$-C$_4$H$_9$+HO$_2$=$\Phi$-C$_4$H$_8$-3+H$_2$O$_2$ | $4.0\times10^{11}$ | 0.0 | 15.5 | j | (31) |
| $\Phi$-C$_4$H$_9$+HO$_2$=$\Phi$-C$_4$H$_8$-2+H$_2$O$_2$ | $4.0\times10^{11}$ | 0.0 | 15.5 | j | (32) |
| $\Phi$-C$_4$H$_9$+HO$_2$=$\Phi$-C$_4$H$_8$-1+H$_2$O$_2$ | $6.0\times10^{11}$ | 0.0 | 17.0 | j | (33) |
| $\Phi$-C$_4$H$_9$+CH$_3$=$\Phi$-C$_4$H$_8$X+CH$_4$ | $1.0\times10^{11}$ | 0.0 | 7.3 | i | (34) |
| $\Phi$-C$_4$H$_9$+CH$_3$=$\Phi$-C$_4$H$_8$-3+CH$_4$ | $2.0\times10^{11}$ | 0.0 | 9.6 | j | (35) |
| $\Phi$-C$_4$H$_9$+CH$_3$=$\Phi$-C$_4$H$_8$-2+CH$_4$ | $2.0\times10^{11}$ | 0.0 | 9.6 | j | (36) |
| $\Phi$-C$_4$H$_9$+CH$_3$=$\Phi$-C$_4$H$_8$-1+CH$_4$ | $3.0\times10^{-1}$ | 4.0 | 8.2 | j | (37) |
| $\Phi$-C$_4$H$_9$+C$_6$H$_5$#=$\Phi$-C$_4$H$_8$X+benzene | $5.3\times10^{13}$ | 0.0 | 12.0 | k | (38) |
| $\Phi$-C$_4$H$_9$+benzyl=$\Phi$-C$_4$H$_8$X+toluene | $1.1\times10^{12}$ | 0.0 | 13.1 | l | (39) |
| $\Phi$-C$_4$H$_9$+C$_6$H$_5$O#=$\Phi$-C$_4$H$_8$X+phenol | $1.1\times10^{12}$ | 0.0 | 13.1 | l | (40) |
| $\Phi$-C$_4$H$_9$+C$_5$H$_5$#=$\Phi$-C$_4$H$_8$X+C$_5$H$_6$# | $1.1\times10^{12}$ | 0.0 | 13.1 | l | (41) |

## Reactions of phenylbutyl radicals (Φ-C₄H₈)

*Isomerisations*

| | | | | | |
|---|---|---|---|---|---|
| Φ-C₄H₈-3=Φ-C₄H₈-1 | $5.1\times10^{9}$ | 1.0 | 39.5 | m | (42) |
| Φ-C₄H₈-3=Φ-C₄H₈X | $1.9\times10^{10}$ | 1.0 | 34.2 | m | (43) |
| Φ-C₄H₈-3=Φ-C₄H₈-2 | $1.9\times10^{10}$ | 1.0 | 38.7 | m | (44) |
| Φ-C₄H₈-2=Φ-C₄H₈X | $3.4\times10^{9}$ | 1.0 | 32.5 | m | (45) |
| Φ-C₄H₈-1=Φ-C₄H₈X | $5.8\times10^{8}$ | 1.0 | 12.8 | m | (46) |

*Decompositions by β-scission*

| | | | | | |
|---|---|---|---|---|---|
| Φ-C₄H₈X=styrene+C₂H₅ | $1.3\times10^{13}$ | 0.0 | 35.9 | n | (47) |
| Φ-C₄H₈X=Φ-C₄H₈-3+H | $3.0\times10^{13}$ | 0.0 | 50.5 | o | (48) |
| Φ-C₄H₈-3=C₆H₅#+C₄H₈Y | $2.0\times10^{13}$ | 0.0 | 34.5 | p | (49) |
| Φ-C₄H₈-3=Φ-C₃H₅Z+CH₃ | $2.0\times10^{13}$ | 0.0 | 32.5 | o | (50) |
| Φ-C₄H₈-3=Φ-C₄H₇-3+H | $3.2\times10^{13}$ | 0.0 | 34.8 | p | (51) |
| Φ-C₄H₈-3=Φ-C₄H₇-2+H | $3.0\times10^{13}$ | 0.0 | 38.0 | p | (52) |
| Φ-C₄H₈-2=benzyl+C₃H₆ | $3.3\times10^{13}$ | 0.0 | 22.5 | p | (53) |
| Φ-C₄H₈-2=Φ-C₄H₇-2+H | $3.0\times10^{13}$ | 0.0 | 38.0 | p | (54) |
| Φ-C₄H₈-2=Φ-C₄H₇-1+H | $3.0\times10^{13}$ | 0.0 | 39.0 | p | (55) |
| Φ-C₄H₈-1=Φ-C₂H₄-1+C₂H₄ | $2.0\times10^{13}$ | 0.0 | 28.7 | j | (56) |
| Φ-C₄H₈-1=Φ-C₄H₇-1+H | $3.0\times10^{13}$ | 0.0 | 38.0 | p | (57) |

*Oxidations*

| | | | | | |
|---|---|---|---|---|---|
| Φ-C₄H₈X+O₂=Φ-C₄H₇-3+HO₂ | $1.6\times10^{12}$ | 0.0 | 15.2 | q | (58) |
| Φ-C₄H₈-3+O₂=Φ-C₄H₇-3+HO₂ | $2.6\times10^{11}$ | 0.0 | 2.5 | q | (59) |
| Φ-C₄H₈-3+O₂=Φ-C₄H₇-2+HO₂ | $1.6\times10^{12}$ | 0.0 | 5.0 | j | (60) |
| Φ-C₄H₈-2+O₂=Φ-C₄H₇-2+HO₂ | $1.6\times10^{12}$ | 0.0 | 5.0 | j | (61) |
| Φ-C₄H₈-2+O₂=Φ-C₄H₇-1+HO₂ | $6.9\times10^{11}$ | 0.0 | 5.0 | j | (62) |
| Φ-C₄H₈-1+O₂=Φ-C₄H₇-1+HO₂ | $1.6\times10^{12}$ | 0.0 | 5.0 | j | (63) |

*Combinations and disproportionations*

| | | | | | |
|---|---|---|---|---|---|
| Φ-C₄H₈X+HO₂=>Φ-C₄H₈O+OH | $5.0\times10^{12}$ | 0.0 | 0.0 | r | (64) |
| Φ-C₄H₈X+CH₃=Φ-C₅H₁₁ | $1.5\times10^{13}$ | 0.0 | 0.0 | b | (65) |
| Φ-C₄H₈X+benzyl=toluene+Φ-C₄H₇-3 | $1.5\times10^{13}$ | 0.0 | 0.0 | b | (66) |
| Φ-C₄H₈X+aC₃H₅=C₃H₆+Φ-C₄H₇-3 | $1.5\times10^{13}$ | 0.0 | 0.0 | b | (67) |

## Reactions of phenylpropyl radical (Φ-C₃H₆-1)

*Isomerisations-decomposition*

| | | | | | |
|---|---|---|---|---|---|
| Φ-C₃H₆-1=>styrene+CH₃ | $3.4\times10^{9}$ | 1.0 | 32.5 | m, s | (68) |

*Cyclisation and reverse internal ipso-addition*

| | | | | | |
|---|---|---|---|---|---|
| Φ-C₃H₆-1=>indane+H | $1.4\times10^{11}$ | 1.0 | 16.2 | t | (69) |
| indane+H=>Φ-C₃H₆-1 | $5.8\times10^{13}$ | 0.0 | 8.1 | e | (-69) |

*Decompositions by β-scission*

| | | | | | |
|---|---|---|---|---|---|
| Φ-C₃H₆-1=benzyl+C₂H₄ | $3.3\times10^{13}$ | 0.0 | 22.5 | p | (70) |
| Φ-C₃H₆-1=Φ-C₃H₅Z+H | $3.0\times10^{13}$ | 0.0 | 38.0 | p | (71) |

*Oxidations*

| | | | | | |
|---|---|---|---|---|---|
| Φ-C₃H₆-1+O₂=Φ-C₃H₅Z+HO₂ | $1.6\times10^{12}$ | 0.0 | 5.0 | j | (72) |

*Combinations*

| | | | | | |
|---|---|---|---|---|---|
| Φ-C₃H₆-1+H=Φ-C₃H₇ | $1.0\times10^{14}$ | 0.0 | 0.0 | c | (73) |

## Reactions of 1-butyl radicals (C₄H₉)

*Isomerisations followed by rapid decomposition*

| | | | | | |
|---|---|---|---|---|---|
| C₄H₉=>CH₃+C₃H₆ | $3.3\times10^{9}$ | 1.0 | 37.0 | m | (74) |

*Decompositions by β-scission*

| | | | | | |
|---|---|---|---|---|---|
| C₄H₉=C₂H₄+C₂H₅ | $2.0\times10^{13}$ | 0.0 | 28.7 | j | (75) |
| C₄H₉=C₄H₈Y+H | $3.0\times10^{13}$ | 0.0 | 38.0 | j | (76) |

*Oxidations*

| | | | | | |
|---|---|---|---|---|---|
| C₄H₉+O₂=C₄H₈Y+HO₂ | $1.6\times10^{12}$ | 0.0 | 5.0 | j | (77) |

*Combinations*

| | | | | | |
|---|---|---|---|---|---|
| C₄H₉+H=C₄H₁₀ | $1.0\times10^{14}$ | 0.0 | 0.0 | c | (78) |

## Reactions of butylphenoxy radical (OC₆H₄#C₄H₉)

*CO Eliminations followed by rapid decompositions*

| | | | | | |
|---|---|---|---|---|---|
| OC₆H₄#C₄H₉=>C₅H₅#+C₄H₈Y+CO | $2.5\times10^{11}$ | 0.0 | 43.8 | u | (79) |
| $OC_6H_4\#C_4H_9 \Rightarrow benzene+C_3H_6+CO+H$ | $2.5\times10^{11}$ | 0.0 | 43.8 | u | (80) |

## Reactions of propylbenzylalcoxy radical ($\Phi\text{-}C_4H_8O$)

*Isomerisations followed by decompositions*

| | | | | | |
|---|---|---|---|---|---|
| $\Phi\text{-}C_4H_8O \Rightarrow Styrene+OH+C_2H_4$ | $1.5\times10^8$ | 1.0 | 8.6 | m | (81) |
| $\Phi\text{-}C_4H_8O \Rightarrow \Phi\text{-}CHOH+C_3H_6$ | $5.7\times10^8$ | 1.0 | 12.1 | m | (82) |

*Decompositions by β-scission*

| | | | | | |
|---|---|---|---|---|---|
| $\Phi\text{-}C_4H_8O = \Phi\text{-}CHO+nC_3H_7$ | $2.0\times10^{13}$ | 0.0 | 15.0 | v | (83) |
| $\Phi\text{-}C_4H_8O = C_6H_5\#+C_3H_7CHO$ | $2.0\times10^{13}$ | 0.0 | 34.5 | p | (84) |

# SECONDARY MECHANISM

## Reactions of phenylbutenes and derived radicals ($\Phi\text{-}C_4H_7$)

*Bimolecular initiation*

| | | | | | |
|---|---|---|---|---|---|
| $\Phi\text{-}C_4H_7\text{-}1+O_2 = \Phi\text{-}C_4H_6X+HO_2$ | $1.4\times10^{12}$ | 0.0 | 35.1 | d | (85) |
| $\Phi\text{-}C_4H_7\text{-}1+O_2 = \Phi\text{-}C_4H_6X(1\text{-}3)+HO_2$ | $1.4\times10^{12}$ | 0.0 | 35.1 | d | (86) |
| $\Phi\text{-}C_4H_7\text{-}2+O_2 = \Phi\text{-}C_4H_6X(1\text{-}3)+HO_2$ | $1.4\times10^{12}$ | 0.0 | 35.1 | d | (87) |
| $\Phi\text{-}C_4H_7\text{-}2+O_2 = \Phi\text{-}C_4H_6Y(2\text{-}4)+HO_2$ | $1.4\times10^{12}$ | 0.0 | 35.1 | d | (88) |
| $\Phi\text{-}C_4H_7\text{-}3+O_2 = \Phi\text{-}C_4H_6Y(2\text{-}4)+HO_2$ | $1.4\times10^{12}$ | 0.0 | 35.1 | d | (89) |
| $\Phi\text{-}C_4H_7\text{-}3+O_2 = \Phi\text{-}C_4H_6\text{-}1+HO_2$ | $2.1\times10^{13}$ | 0.0 | 52.3 | d | (90) |

*Ipso-additions*

| | | | | | |
|---|---|---|---|---|---|
| $\Phi\text{-}C_4H_7\text{-}1+H=benzene+C_4H_7Y$ | $5.8\times10^{13}$ | 0.0 | 8.1 | e | (91) |
| $\Phi\text{-}C_4H_7\text{-}1+CH_3=toluene+C_4H_7Y$ | $1.2\times10^{12}$ | 0.0 | 15.9 | f | (92) |
| $\Phi\text{-}C_4H_7\text{-}2+H=benzene+C_4H_7Y$ | $5.8\times10^{13}$ | 0.0 | 8.1 | e | (93) |
| $\Phi\text{-}C_4H_7\text{-}2+CH_3=toluene+C_4H_7Y$ | $1.2\times10^{12}$ | 0.0 | 15.9 | f | (94) |
| $\Phi\text{-}C_4H_7\text{-}3+H=benzene+C_4H_7Y$ | $5.8\times10^{13}$ | 0.0 | 8.1 | e | (95) |
| $\Phi\text{-}C_4H_7\text{-}3+CH_3=toluene+C_4H_7Y$ | $5.0\times10^{12}$ | 0.0 | 15.9 | f | (96) |

*Additions to the double bond followed by rapid decomposition*

| | | | | | |
|---|---|---|---|---|---|
| $\Phi\text{-}C_4H_7\text{-}1+O \Rightarrow \Phi\text{-}C_2H_4\text{-}1+CH_2CHO$ | $6.4\times10^4$ | 2.6 | -1.1 | i | (97) |
| $\Phi\text{-}C_4H_7\text{-}1+O \Rightarrow benzyl+HCHO+C_2H_3V$ | $6.4\times10^4$ | 2.6 | -1.1 | i | (98) |
| $\Phi\text{-}C_4H_7\text{-}1+OH \Rightarrow \Phi\text{-}C_3H_6\text{-}1+HCHO$ | $1.4\times10^{12}$ | 0.0 | -1.0 | i | (99) |
| $\Phi\text{-}C_4H_7\text{-}1+OH \Rightarrow \Phi\text{-}C_2H_4\text{-}1+CH_3CHO$ | $1.4\times10^{12}$ | 0.0 | -1.0 | i | (100) |
| $\Phi\text{-}C_4H_7\text{-}1+CH_3 \Rightarrow \Phi\text{-}C_2H_4\text{-}1+C_3H_6Y$ | $1.7\times10^{11}$ | 0.0 | 7.4 | i | (101) |
| $\Phi\text{-}C_4H_7\text{-}2+O \Rightarrow \Phi\text{-}C_2H_4\text{-}1+CH_2CHO$ | $6.4\times10^4$ | 2.6 | -1.1 | i | (102) |
| $\Phi\text{-}C_4H_7\text{-}2+O \Rightarrow benzyl+C_2H_3CHO+H$ | $6.4\times10^4$ | 2.6 | -1.1 | i | (103) |
| $\Phi\text{-}C_4H_7\text{-}2+OH \Rightarrow \Phi\text{-}C_2H_4\text{-}1+CH_3CHO$ | $1.4\times10^{12}$ | 0.0 | -1.0 | i | (104) |
| $\Phi\text{-}C_4H_7\text{-}2+OH \Rightarrow benzyl+C_2H_5CHO$ | $1.4\times10^{12}$ | 0.0 | -1.0 | i | (105) |
| $\Phi\text{-}C_4H_7\text{-}2+CH_3 \Rightarrow benzyl+C_4H_8Y$ | $1.7\times10^{11}$ | 0.0 | 7.4 | i | (106) |
| $\Phi\text{-}C_4H_7\text{-}3+O \Rightarrow C_6H_5\#+C_2H_3CHO+CH_3$ | $6.4\times10^4$ | 2.6 | -1.1 | i | (107) |
| $\Phi\text{-}C_4H_7\text{-}3+OH \Rightarrow \Phi\text{-}CHO+n\text{-}C_3H_7$ | $1.4\times10^{12}$ | 0.0 | -1.0 | i | (108) |
| $\Phi\text{-}C_4H_7\text{-}3+OH \Rightarrow benzyl+C_2H_5CHO$ | $1.4\times10^{12}$ | 0.0 | -1.0 | i | (109) |
| $\Phi\text{-}C_4H_7\text{-}3+CH_3 \Rightarrow \Phi\text{-}C_3H_5Z+C_2H_5$ | $1.7\times10^{11}$ | 0.0 | 7.4 | i | (110) |

*Metatheses*

| | | | | | |
|---|---|---|---|---|---|
| $\Phi\text{-}C_4H_7\text{-}1+O = \Phi\text{-}C_4H_6X+OH$ | $8.8\times10^{10}$ | 0.7 | 3.3 | i | (111) |
| $\Phi\text{-}C_4H_7\text{-}1+O = \Phi\text{-}C_4H_6X(1\text{-}3)+OH$ | $8.8\times10^{10}$ | 0.7 | 3.3 | i | (112) |
| $\Phi\text{-}C_4H_7\text{-}1+H = \Phi\text{-}C_4H_6X+H_2$ | $5.4\times10^4$ | 2.5 | -1.9 | i | (113) |
| $\Phi\text{-}C_4H_7\text{-}1+H = \Phi\text{-}C_4H_6X(1\text{-}3)+H_2$ | $5.4\times10^4$ | 2.5 | -1.9 | i | (114) |
| $\Phi\text{-}C_4H_7\text{-}1+OH = \Phi\text{-}C_4H_6X+H_2O$ | $3.0\times10^6$ | 2.0 | -1.5 | i | (115) |
| $\Phi\text{-}C_4H_7\text{-}1+OH = \Phi\text{-}C_4H_6X(1\text{-}3)+H_2O$ | $3.0\times10^6$ | 2.0 | -1.5 | i | (116) |
| $\Phi\text{-}C_4H_7\text{-}1+HO_2 = \Phi\text{-}C_4H_6X+H_2O_2$ | $6.4\times10^3$ | 2.6 | 12.4 | i | (117) |
| $\Phi\text{-}C_4H_7\text{-}1+HO_2 = \Phi\text{-}C_4H_6X(1\text{-}3)+H_2O_2$ | $6.4\times10^3$ | 2.6 | 12.4 | i | (118) |
| $\Phi\text{-}C_4H_7\text{-}1+CH_3 = \Phi\text{-}C_4H_6X+CH_4$ | $1.0\times10^{11}$ | 0.0 | 7.3 | i | (119) |
| $\Phi\text{-}C_4H_7\text{-}1+CH_3 = \Phi\text{-}C_4H_6X(1\text{-}3)+CH_4$ | $1.0\times10^{11}$ | 0.0 | 7.3 | i | (120) |
| $\Phi\text{-}C_4H_7\text{-}1+benzyl = \Phi\text{-}C_4H_6X+toluene$ | $1.1\times10^{12}$ | 0.0 | 13.1 | l | (121) |
| $\Phi\text{-}C_4H_7\text{-}1+benzyl = \Phi\text{-}C_4H_6X(1\text{-}3)+toluene$ | $1.1\times10^{12}$ | 0.0 | 13.1 | l | (122) |
| $\Phi\text{-}C_4H_7\text{-}2+O = \Phi\text{-}C_4H_6Y(2\text{-}4)+OH$ | $8.8\times10^{10}$ | 0.7 | 3.3 | i | (123) |
| $\Phi\text{-}C_4H_7\text{-}2+O = \Phi\text{-}C_4H_6X(1\text{-}3)+OH$ | $1.7\times10^{11}$ | 0.7 | 5.9 | i | (124) |
| $\Phi\text{-}C_4H_7\text{-}2+H = \Phi\text{-}C_4H_6Y(2\text{-}4)+H_2$ | $5.4\times10^4$ | 2.5 | -1.9 | i | (125) |
| $\Phi\text{-}C_4H_7\text{-}2+H = \Phi\text{-}C_4H_6X(1\text{-}3)+H_2$ | $1.7\times10^5$ | 2.5 | 2.5 | i | (126) |
| $\Phi\text{-}C_4H_7\text{-}2+OH = \Phi\text{-}C_4H_6Y(2\text{-}4)+H_2O$ | $3.0\times10^6$ | 2.0 | -1.5 | i | (127) |
| $\Phi$-$C_4H_7$-2+OH=$\Phi$-$C_4H_6X(1\text{-}3)$+$H_2O$ | $3.0\times10^6$ | 2.0 | -0.3 | i | (128) |
| $\Phi$-$C_4H_7$-2+$HO_2$=$\Phi$-$C_4H_6Y(2\text{-}4)$+$H_2O_2$ | $6.4\times10^3$ | 2.6 | 12.4 | i | (129) |
| $\Phi$-$C_4H_7$-2+$HO_2$=$\Phi$-$C_4H_6X(1\text{-}3)$+$H_2O_2$ | $9.6\times10^3$ | 2.6 | 13.9 | i | (130) |
| $\Phi$-$C_4H_7$-2+$CH_3$=$\Phi$-$C_4H_6Y(2\text{-}4)$+$CH_4$ | $1.0\times10^{11}$ | 0.0 | 7.3 | i | (131) |
| $\Phi$-$C_4H_7$-2+$CH_3$=$\Phi$-$C_4H_6X(1\text{-}3)$+$CH_4$ | 2.2 | 3.5 | 5.7 | i | (132) |
| $\Phi$-$C_4H_7$-2+benzyl=$\Phi$-$C_4H_6Y(2\text{-}4)$+toluene | $1.1\times10^{12}$ | 0.0 | 13.1 | l | (133) |
| $\Phi$-$C_4H_7$-2+benzyl=$\Phi$-$C_4H_6X(1\text{-}3)$+toluene | $1.1\times10^{12}$ | 0.0 | 13.1 | l | (134) |
| $\Phi$-$C_4H_7$-3+O=$\Phi$-$C_4H_6Y(2\text{-}4)$+OH | $8.8\times10^{10}$ | 0.7 | 3.3 | i | (135) |
| $\Phi$-$C_4H_7$-3+O=$\Phi$-$C_4H_6$-1+OH | $5.1\times10^{13}$ | 0.7 | 7.9 | i | (136) |
| $\Phi$-$C_4H_7$-3+H=$\Phi$-$C_4H_6Y(2\text{-}4)$+$H_2$ | $5.4\times10^4$ | 2.5 | -1.9 | i | (137) |
| $\Phi$-$C_4H_7$-3+H=$\Phi$-$C_4H_6$-1+$H_2$ | $2.9\times10^7$ | 2.0 | 7.7 | i | (138) |
| $\Phi$-$C_4H_7$-3+OH=$\Phi$-$C_4H_6Y(2\text{-}4)$+$H_2O$ | $3.0\times10^6$ | 2.0 | -1.5 | i | (139) |
| $\Phi$-$C_4H_7$-3+OH=$\Phi$-$C_4H_6$-1+$H_2O$ | $2.7\times10^6$ | 2.0 | 0.5 | i | (140) |
| $\Phi$-$C_4H_7$-3+$HO_2$=$\Phi$-$C_4H_6Y(2\text{-}4)$+$H_2O_2$ | $6.4\times10^3$ | 2.6 | 12.4 | i | (141) |
| $\Phi$-$C_4H_7$-3+$HO_2$=$\Phi$-$C_4H_6$-1+$H_2O_2$ | $6.0\times10^{11}$ | 0.0 | 17.0 | i | (142) |
| $\Phi$-$C_4H_7$-3+$CH_3$=$\Phi$-$C_4H_6Y(2\text{-}4)$+$CH_4$ | $1.0\times10^{11}$ | 0.0 | 7.3 | i | (143) |
| $\Phi$-$C_4H_7$-3+$CH_3$=$\Phi$-$C_4H_6$-1+$CH_4$ | $3.0\times10^{-1}$ | 4.0 | 8.2 | i | (144) |
| $\Phi$-$C_4H_7$-3+benzyl=$\Phi$-$C_4H_6Y(2\text{-}4)$+toluene | $1.1\times10^{12}$ | 0.0 | 13.1 | l | (145) |

*Decompositions of the obtained radicals by $\beta$-scission*

| | | | | | |
|---|---|---|---|---|---|
| $\Phi$-$C_4H_6X$=styrene+$C_2H_3$ | $2.0\times10^{13}$ | 0.0 | 44.2 | w | (146) |
| $\Phi$-$C4H6Y(2\text{-}4)$=$\Phi$-$C_4H_5$+H | $3.0\times10^{13}$ | 0.0 | 51.5 | p | (147) |
| $\Phi$-$C_4H_6X(1\text{-}3)$=$C_6H_5$#+$C_4H_6(1\text{-}3)$ | $2.0\times10^{13}$ | 0.0 | 44.2 | w | (148) |
| $\Phi$-$C_4H_6X$=$\Phi$-$C_4H_5$+H | $3.0\times10^{13}$ | 0.0 | 43.3 | w | (149) |
| $\Phi$-$C_4H_6X(1\text{-}3)$=$\Phi$-$C_4H_5$+H | $3.0\times10^{13}$ | 0.0 | 43.3 | w | (150) |
| $\Phi$-$C_4H_6$-1=$\Phi$-$C_4H_5$+H | $3.0\times10^{13}$ | 0.0 | 39.0 | p | (151) |

*Oxidations of the obtained radicals*

| | | | | | |
|---|---|---|---|---|---|
| $\Phi$-$C_4H_6X$+$O_2$=$\Phi$-$C_4H_5$+$HO_2$ | $7.9\times10^{11}$ | 0.0 | 15.2 | q | (152) |
| $\Phi$-$C_4H_6Y(2\text{-}4)$+$O_2$=$\Phi$-$C_4H_5$+$HO_2$ | $6.9\times10^{11}$ | 0.0 | 15.2 | q | (153) |
| $\Phi$-$C_4H_6X(1\text{-}3)$+$O_2$=$\Phi$-$C_4H_5$+$HO_2$ | $7.9\times10^{11}$ | 0.0 | 15.2 | q | (154) |
| $\Phi$-$C_4H_6$-1+$O_2$=$\Phi$-$C_4H_5$+$HO_2$ | $2.6\times10^{11}$ | 0.0 | 2.5 | q | (155) |

*Cyclisations of the obtained radicals*

| | | | | | |
|---|---|---|---|---|---|
| $\Phi$-$C_4H_6Y(2\text{-}4)$=$C_{10}H_{10}$#+H | $1.4\times10^{11}$ | 0.0 | 16.2 | t | (156) |
| $\Phi$-$C_4H_6X(1\text{-}3)$=$C_{10}H_{10}$#+H | $1.4\times10^{11}$ | 0.0 | 16.2 | t | (157) |
| $\Phi$-$C_4H_6$-1=$C_{10}H_{10}$#+H | $1.4\times10^{11}$ | 0.0 | 16.2 | t | (158) |

*Combinations of the obtained radicals with H-atoms*

| | | | | | |
|---|---|---|---|---|---|
| $\Phi$-$C_4H_6X$+H=$\Phi$-$C_4H_7$-1 | $1.0\times10^{14}$ | 0.0 | 0.0 | c | (159) |
| $\Phi$-$C_4H_6Y(2\text{-}4)$+H=$\Phi$-$C_4H_7$-2 | $1.0\times10^{14}$ | 0.0 | 0.0 | c | (160) |
| $\Phi$-$C_4H_6Y(2\text{-}4)$+H=$\Phi$-$C_4H_7$-3 | $1.0\times10^{14}$ | 0.0 | 0.0 | c | (161) |
| $\Phi$-$C_4H_6X(1\text{-}3)$+H=$\Phi$-$C_4H_7$-1 | $1.0\times10^{14}$ | 0.0 | 0.0 | c | (162) |
| $\Phi$-$C_4H_6X(1\text{-}3)$+H=$\Phi$-$C_4H_7$-2 | $1.0\times10^{14}$ | 0.0 | 0.0 | c | (163) |
| $\Phi$-$C_4H_6$-1+H=$\Phi$-$C_4H_7$-3 | $1.0\times10^{14}$ | 0.0 | 0.0 | c | (164) |

## Reactions of allylbenzene ($\Phi$-$C_3H_5Z$) and derived radicals

*Bimolecular initiation*

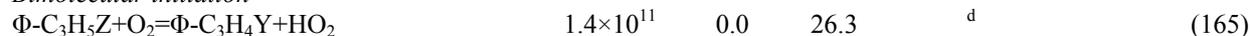

| | | | | | |
|---|---|---|---|---|---|
| $\Phi$-$C_3H_5Z$+$O_2$=$\Phi$-$C_3H_4Y$+$HO_2$ | $1.4\times10^{11}$ | 0.0 | 26.3 | d | (165) |

*Ipso-additions*

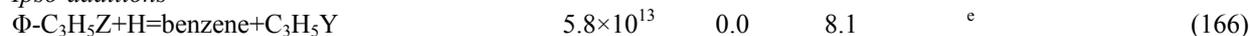
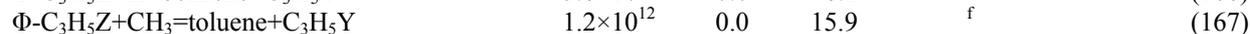

| | | | | | |
|---|---|---|---|---|---|
| $\Phi$-$C_3H_5Z$+H=benzene+$C_3H_5Y$ | $5.8\times10^{13}$ | 0.0 | 8.1 | e | (166) |
| $\Phi$-$C_3H_5Z$+$CH_3$=toluene+$C_3H_5Y$ | $1.2\times10^{12}$ | 0.0 | 15.9 | f | (167) |

*Additions to the double bond followed by rapid decompositions*

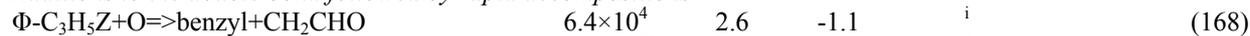
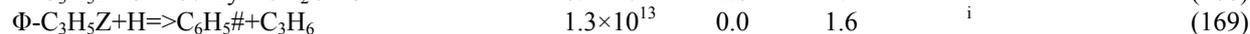
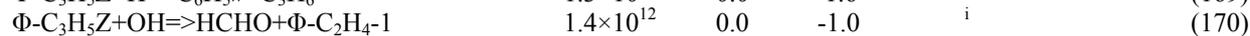

| | | | | | |
|---|---|---|---|---|---|
| $\Phi$-$C_3H_5Z$+O=>benzyl+$CH_2CHO$ | $6.4\times10^4$ | 2.6 | -1.1 | i | (168) |
| $\Phi$-$C_3H_5Z$+H=>$C_6H_5$#+$C_3H_6$ | $1.3\times10^{13}$ | 0.0 | 1.6 | i | (169) |
| $\Phi$-$C_3H_5Z$+OH=>HCHO+$\Phi$-$C_2H_4$-1 | $1.4\times10^{12}$ | 0.0 | -1.0 | i | (170) |
| $\Phi$-$C_3H_5Z$+$CH_3$=>benzyl+$C_3H_6Y$ | $1.7\times10^{11}$ | 0.0 | 7.4 | i | (171) |

*Metatheses*

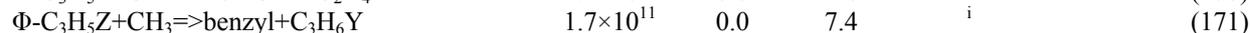
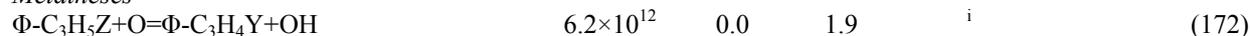
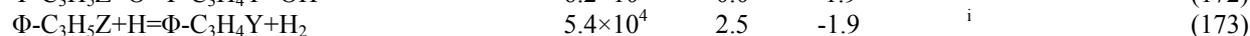
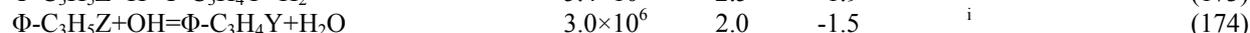
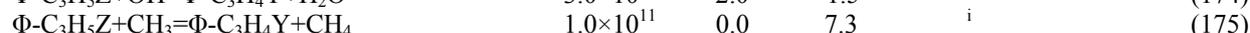
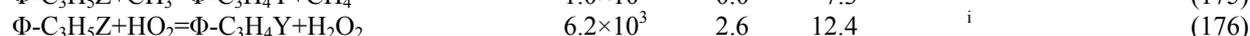
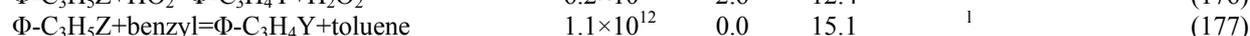

| | | | | | |
|---|---|---|---|---|---|
| $\Phi$-$C_3H_5Z$+O=$\Phi$-$C_3H_4Y$+OH | $6.2\times10^{12}$ | 0.0 | 1.9 | i | (172) |
| $\Phi$-$C_3H_5Z$+H=$\Phi$-$C_3H_4Y$+$H_2$ | $5.4\times10^4$ | 2.5 | -1.9 | i | (173) |
| $\Phi$-$C_3H_5Z$+OH=$\Phi$-$C_3H_4Y$+$H_2O$ | $3.0\times10^6$ | 2.0 | -1.5 | i | (174) |
| $\Phi$-$C_3H_5Z$+$CH_3$=$\Phi$-$C_3H_4Y$+$CH_4$ | $1.0\times10^{11}$ | 0.0 | 7.3 | i | (175) |
| $\Phi$-$C_3H_5Z$+$HO_2$=$\Phi$-$C_3H_4Y$+$H_2O_2$ | $6.2\times10^3$ | 2.6 | 12.4 | i | (176) |
| $\Phi$-$C_3H_5Z$+benzyl=$\Phi$-$C_3H_4Y$+toluene | $1.1\times10^{12}$ | 0.0 | 15.1 | l | (177) |

*Decompositions and combinations of the obtained radical*

| | | | | | |
|---|---|---|---|---|---|
| $\Phi$-$C_3H_4Y$=$\Phi$-$C_3H_3$+H | $1.5\times10^{13}$ | 0.0 | 46.0 | [p] | (178) |
| $\Phi$-$C_3H_4Y$+H=$\Phi$-$C_3H_5Z$ | $1.0\times10^{14}$ | 0.0 | 0.0 | [e] | (179) |
| $\Phi$-$C_3H_4Y$+$CH_3$=$\Phi$-$C_4H_7$-1 | $1.5\times10^{13}$ | 0.0 | 0.0 | [b] | (180) |
| $\Phi$-$C_3H_4Y$+$HO_2$=OH+$C_2H_3V$+$\Phi$-CHO | $5.0\times10^{12}$ | 0.0 | 0.0 | [r] | (181) |

*Cyclisation of the obtained radical*

| | | | | | |
|---|---|---|---|---|---|
| $\Phi$-$C_3H_4Y$=indene+H | $1.4\times10^{11}$ | 0.0 | 16.2 | [t] | (182) |

## Reactions of propylbenzene ($\Phi$-$C_3H_7$)

*Unimolecular initiation*

| | | | | | |
|---|---|---|---|---|---|
| $\Phi$-$C_2H_4$-1+$CH_3$=$\Phi$-$C_3H_7$ | $1.5\times10^{13}$ | 0.0 | 0.0 | [b] | (183) |
| benzyl+$C_2H_5$=$\Phi$-$C_3H_7$ | $1.5\times10^{13}$ | 0.0 | 0.0 | [b] | (184) |
| $C_6H_5$#+$nC_3H_7$=$\Phi$-$C_3H_7$ | $1.5\times10^{13}$ | 0.0 | 0.0 | [b] | (185) |

*Bimolecular initiations*

| | | | | | |
|---|---|---|---|---|---|
| $\Phi$-$C_3H_7$+$O_2$=>styrene+$CH_3$+$HO_2$ | $1.4\times10^{12}$ | 0.0 | 35.1 | [d] | (186) |
| $\Phi$-$C_3H_7$+$O_2$=$\Phi$-$C_3H_6$-1+$HO_2$ | $2.1\times10^{13}$ | 0.0 | 52.3 | [d] | (187) |

*Ipso-addition*

| | | | | | |
|---|---|---|---|---|---|
| $\Phi$-$C_3H_7$+H=benzene+$nC_3H_7$ | $5.8\times10^{13}$ | 0.0 | 8.1 | [e] | (188) |
| $\Phi$-$C_3H_7$+$CH_3$=toluene+$nC_3H_7$ | $1.2\times10^{12}$ | 0.0 | 15.9 | [f] | (189) |

*Metatheses*

| | | | | | |
|---|---|---|---|---|---|
| $\Phi$-$C_3H_7$+O=>$CH_3$+styrene+OH | $8.8\times10^{10}$ | 0.7 | 3.3 | [i] | (190) |
| $\Phi$-$C_3H_7$+O=$\Phi$-$C_3H_6$-1+OH | $5.1\times10^{13}$ | 0.0 | 7.9 | [j] | (191) |
| $\Phi$-$C_3H_7$+H=>$CH_3$+styrene+$H_2$ | $5.4\times10^{4}$ | 2.5 | -1.9 | [i] | (192) |
| $\Phi$-$C_3H_7$+H=$\Phi$-$C_3H_6$-1+$H_2$ | $2.8\times10^{7}$ | 2.0 | 7.7 | [j] | (193) |
| $\Phi$-$C_3H_7$+OH=>$CH_3$+styrene+$H_2O$ | $3.0\times10^{6}$ | 2.0 | -1.5 | [i] | (194) |
| $\Phi$-$C_3H_7$+OH=$\Phi$-$C_3H_6$-1+$H_2O$ | $2.7\times10^{6}$ | 2.0 | 0.5 | [j] | (195) |
| $\Phi$-$C_3H_7$+$HO_2$=>$CH_3$+styrene+$H_2O_2$ | $6.4\times10^{3}$ | 2.6 | 12.4 | [i] | (196) |
| $\Phi$-$C_3H_7$+$HO_2$=$\Phi$-$C_3H_6$-1+$H_2O_2$ | $6.0\times10^{11}$ | 0.0 | 17.0 | [j] | (197) |
| $\Phi$-$C_3H_7$+$CH_3$=>$CH_3$+styrene+$CH_4$ | $1.0\times10^{11}$ | 0.0 | 7.3 | [i] | (198) |
| $\Phi$-$C_3H_7$+$CH_3$=$\Phi$-$C_3H_6$-1+$CH_4$ | $3.0\times10^{1}$ | 4.0 | 8.2 | [j] | (199) |
| $\Phi$-$C_3H_7$+benzyl=>$CH_3$+styrene+toluene | $1.1\times10^{12}$ | 0.0 | 13.1 | [l] | (200) |

## Reactions of 2-phenyl-n-pentane ($\Phi$-$C_5H_{11}$)

*H-abstractions followed by β-scission decomposition*

| | | | | | |
|---|---|---|---|---|---|
| $\Phi$-$C_5H_{11}$+H=>$H_2$+styrene+$nC_3H_7$ | $3.9\times10^{7}$ | 2.0 | 7.7 | [j] | (201) |
| $\Phi$-$C_5H_{11}$+H=>$H_2$+$\Phi$-$C_2H_4X$+$C_3H_6Y$ | $1.8\times10^{7}$ | 2.0 | 5.0 | [j] | (202) |
| $\Phi$-$C_5H_{11}$+OH=>$H_2O$+styrene+$nC_3H_7$ | $5.3\times10^{6}$ | 2.0 | 0.5 | [j] | (203) |
| $\Phi$-$C_5H_{11}$+OH=>$H_2O$+$\Phi$-$C_2H_4X$+$C_3H_6Y$ | $5.2\times10^{6}$ | 2.0 | -0.8 | [j] | (204) |

## Reactions of indane

*Bimolecular initiations followed by rapid decompositions*

| | | | | | |
|---|---|---|---|---|---|
| indane+$O_2$=>$HO_2$+indene+H | $2.8\times10^{12}$ | 0.0 | 35.1 | [d] | (205) |
| indane+$O_2$=>$HO_2$+indene+H | $8.0\times10^{12}$ | 0.0 | 49.9 | [d] | (206) |

*Metatheses followed by rapid decompositions*

| | | | | | |
|---|---|---|---|---|---|
| indane+O=>indene+H+OH | $1.8\times10^{11}$ | 0.7 | 3.3 | [i] | (207) |
| indane+O=>indene+H+OH | $2.6\times10^{13}$ | 0.0 | 5.2 | [i] | (208) |
| indane+H=>indene+H+$H_2$ | $1.1\times10^{5}$ | 2.5 | -1.9 | [i] | (209) |
| indane+H=>indene+H+$H_2$ | $9.0\times10^{6}$ | 2.0 | 5.0 | [i] | (210) |
| indane+OH=>indene+H+$H_2O$ | $6.0\times10^{6}$ | 2.0 | -1.5 | [i] | (211) |
| indane+OH=>indene+H+$H_2O$ | $2.6\times10^{6}$ | 2.0 | -0.8 | [i] | (212) |
| indane+$CH_3$=>indene+H+$CH_4$ | $6.2\times10^{11}$ | 0.0 | 5.5 | [i] | (213) |
| indane+$CH_3$=>indene+H+$CH_4$ | $2.0\times10^{11}$ | 0.0 | 9.6 | [i] | (214) |
| indane+$HO_2$=>$H_2O_2$+indene+H | $1.3\times10^{4}$ | 2.6 | 12.4 | [i] | (215) |
| indane+$HO_2$=>$H_2O_2$+indene+H | $4.0\times10^{11}$ | 0.0 | 15.5 | [i] | (216) |

## Reactions of butadienylbenzene ($\Phi$-$C_4H_5$)

*Bimolecular initiations followed by cyclizations*

| | | | | | |
|---|---|---|---|---|---|
| $\Phi$-$C_4H_5$+$O_2$=>$HO_2$+naphthalene+H | $2.0\times10^{13}$ | 0.0 | 57.6 | [d] | (217) |
| $\Phi$-$C_4H_5$+$O_2$=>$HO_2$+naphthalene+H | $4.0\times10^{12}$ | 0.0 | 55.6 | [d] | (218) |

*Metatheses followed by cyclizations*

| | | | | | |
|---|---|---|---|---|---|
| $\Phi$-$C_4H_5$+O=>naphthalene+H+OH | $1.2\times10^{11}$ | 0.7 | 9.0 | i | (219) |
| $\Phi$-$C_4H_5$+O=>naphthalene+H+OH | $1.2\times10^{11}$ | 0.7 | 7.6 | i | (220) |
| $\Phi$-$C_4H_5$+H=>naphthalene+H+$H_2$ | $8.2\times10^5$ | 2.5 | 12.3 | i | (221) |
| $\Phi$-$C_4H_5$+H=>naphthalene+H+$H_2$ | $8.2\times10^5$ | 2.5 | 9.8 | i | (222) |
| $\Phi$-$C_4H_5$+OH=>naphthalene+H+$H_2O$ | $2.2\times10^6$ | 2.0 | 2.8 | i | (223) |
| $\Phi$-$C_4H_5$+OH=>naphthalene+H+$H_2O$ | $2.2\times10^6$ | 2.0 | 1.5 | i | (224) |
| $\Phi$-$C_4H_5$+$CH_3$=>naphthalene+H+$CH_4$ | 1.4 | 3.5 | 12.9 | i | (225) |
| $\Phi$-$C_4H_5$+$CH_3$=>naphthalene+H+$CH_4$ | 2.0 | 3.5 | 11.7 | i | (226) |

## Reactions of bicyclic isomers of $C_{10}H_{10}$

*Molecular hydrogen elimination*

| | | | | | |
|---|---|---|---|---|---|
| $C_{10}H_{10}$#=naphthalene+$H_2$ | $2.5\times10^{13}$ | 0.0 | 59.0 | x | (227) |

*Metatheses followed by $\beta$-scission decompositions*

| | | | | | |
|---|---|---|---|---|---|
| $C_{10}H_{10}$#+O=>naphthalene+H+OH | $1.8\times10^{11}$ | 0.7 | 3.3 | i | (228) |
| $C_{10}H_{10}$#+H=>naphthalene+H+$H_2$ | $1.1\times10^5$ | 2.5 | -1.9 | i | (229) |
| $C_{10}H_{10}$#+OH=>naphthalene+H+$H_2O$ | $6.0\times10^6$ | 2.0 | -1.5 | i | (230) |
| $C_{10}H_{10}$#+CH3=>naphthalene+H+$CH_4$ | $6.2\times10^{11}$ | 0.0 | 5.5 | i | (231) |

## Reactions of indene and derived radicals

*Bimolecular initiation*

| | | | | | |
|---|---|---|---|---|---|
| indene+$O_2$=indenyl+$HO_2$ | $1.4\times10^{12}$ | 0.0 | 35.1 | d | (232) |

*Additions to the double bond*

| | | | | | |
|---|---|---|---|---|---|
| indene+OH=>$C_2H_4$+$\Phi$-CO | $1.4\times10^{12}$ | 0.0 | -1.0 | i | (233) |

*Metatheses*

| | | | | | |
|---|---|---|---|---|---|
| indene+O=indenyl+OH | $8.8\times10^{10}$ | 0.7 | 3.3 | i | (234) |
| indene+H=indenyl+$H_2$ | $5.4\times10^4$ | 2.5 | -1.9 | i | (235) |
| indene+OH=indenyl+$H_2O$ | $3.0\times10^6$ | 2.0 | -1.5 | i | (236) |
| indene+$CH_3$=indenyl+$CH_4$ | $3.1\times10^{11}$ | 0.0 | 5.5 | i | (237) |
| indene+$HO_2$=$H_2O_2$+indenyl | $6.4\times10^3$ | 2.6 | 12.4 | i | (238) |

*Combinations of the obtained radical*

| | | | | | |
|---|---|---|---|---|---|
| indenyl+H=indene | $1.0\times10^{14}$ | 0.0 | 0.0 | c | (239) |
| indenyl+$CH_3$=>$C_{10}H_{10}$# | $1.5\times10^{13}$ | 0.0 | 0.0 | b | (240) |
| indenyl+O=>$\Phi$-$C_2H$+CHO | $3.2\times10^{13}$ | -0.17 | 0.44 | y | (241) |
| indenyl+O=>$\Phi$-CO+$C_2H_2$ | $3.2\times10^{13}$ | -0.17 | 0.44 | y | (242) |
| indenyl+OH=>$\Phi$-$C_2H_2$+CO+H | $1.0\times10^{13}$ | 0.0 | 0.0 | y | (243) |

## Reactions of naphthalene and derived radicals

*Bimolecular initiation*

| | | | | | |
|---|---|---|---|---|---|
| naphthalene+$O_2$=$HO_2$+naphthyl | $8.0\times10^{13}$ | 0.0 | 63.4 | z | (244) |

*Ipso-addition*

| | | | | | |
|---|---|---|---|---|---|
| naphthalene+O=>indenyl+CO+H | $2.7\times10^{13}$ | 0.0 | 3.6 | z | (245) |

*Metatheses*

| | | | | | |
|---|---|---|---|---|---|
| naphthalene+H=naphthyl+$H_2$ | $8.0\times10^8$ | 1.0 | 16.8 | z | (246) |
| naphthalene+O=naphthyl+OH | $2.7\times10^{13}$ | 0.0 | 14.7 | z | (247) |
| naphthalene+OH=naphthyl+$H_2O$ | $2.1\times10^8$ | 1.4 | 1.5 | z | (248) |
| naphthalene+$HO_2$=naphthyl+$H_2O_2$ | $7.3\times10^{12}$ | 0.0 | 28.9 | z | (249) |
| naphthalene+$CH_3$=naphthyl+$CH_4$ | $2.7\times10^{12}$ | 0.0 | 15.0 | z | (250) |

*Reactions of the obtained radical*

| | | | | | |
|---|---|---|---|---|---|
| naphthyl+$O_2$=>indenyl+CO+O | $2.6\times10^{13}$ | 0.0 | 6.1 | z | (251) |
| naphthyl+H=naphthalene | $1.0\times10^{14}$ | 0.0 | 0.0 | z | (252) |
| naphthyl+O=>indenyl+CO | $1.0\times10^{14}$ | 0.0 | 0.0 | z | (253) |
| naphthyl+OH=>indenyl+CO+H | $1.0\times10^{13}$ | 0.0 | 0.0 | z | (254) |
| naphthyl+$HO_2$=>indenyl+CO+OH | $5.0\times10^{12}$ | 0.0 | 0.0 | z | (255) |

---

[a]: Rate constant estimated by analogy with the values proposed by Rao et al [32] for toluene

[b]: Rate constant of this recombination calculated by the modified collision theory at 1200 K using software KINGAS [33].

[c]: Rate constant taken equal to that of the recombination of H atoms with alkyl radicals as proposed by Allara et al [34].

[d]: Pre-exponentiel factor is the value proposed by Ingham et al [35]. Activation energy calculated at 1200 K by using software THERGAS [24].

## TABLE II: NAMES, FORMULAE AND HEATS OF FORMATION FOR THE AROMATIC SPECIES CONTAINING AT LEAST 9 ATOMS OF CARBON INVOLVED IN THE MECHANISM OF TABLE I.

The heats of formation have been calculated by software THERGAS [24] at 300 K in kcal.mol$^{-1}$, apart from biaromatic species for which they have been obtained from Burcat and Ruscic [26].

| Species | Structure | $\Delta H_f$ (300K) | Species | Structure | $\Delta H_f$ (300K) |
|---|---|---|---|---|---|
| Butylbenzene ($\Phi$-C$_4$H$_9$) | | -3.3 | 4-phenyl-1-butene ($\Phi$-C$_4$H$_7$-1) | | 27.0 |
| 4-phenylbut-1-yl radical ($\Phi$-C$_4$H$_8$-1) | | 45.5 | 4-phenyl-2-butene ($\Phi$-C$_4$H$_7$-2) | | 26.1 |
| 4-phenylbut-2-yl radical ($\Phi$-C$_4$H$_8$-2) | | 43.3 | 4-phenyl-3-butene ($\Phi$-C$_4$H$_7$-3) | | 22.6 |
| 4-phenylbut-3-yl radical ($\Phi$-C$_4$H$_8$-3) | | 43.3 | 4-phenyl-1-buten-4-yl radical ($\Phi$-C$_4$H$_6$X) Resonance stabilized | | 60.3 |
| 4-phenylbut-4-yl radical ($\Phi$-C$_4$H$_8$X) Resonance stabilized | | 29.9 | 4-phenyl-2-buten-1-yl radical or 4-phenyl-1-buten-3-yl radical $\Phi$-C$_4$H$_6$X(1-3) Resonance stabilized | | 57.0 58.5 |
| 3-phenyl-1-propyl radical ($\Phi$-C$_3$H$_6$-1) | | 50.8 | | | |
| Butylphenoxy radical (OC$_6$H$_4$#C$_4$H$_9$) | | -9.4 | 4-phenyl-3-buten-2-yl radical or 4-phenyl-2-buten-4-yl radical $\Phi$-C$_4$H$_6$X(2-4) Resonance stabilized | | 53.1 |
| Propylbenzylalcoxy ($\Phi$-C$_4$H$_8$O) | | 4.0 | 4-phenyl-3-buten-1-yl radical ($\Phi$-C$_4$H$_6$-1) | | 71.4 |
| 2-phenyl-n-pentane ($\Phi$-C$_5$H$_{11}$) | | -9.0 | 4-phenyl-butadiene ($\Phi$-C$_4$H$_5$) | | 48.8 |
| Propylbenzene ($\Phi$-C$_3$H$_7$) | | 2.0 | Allylbenzene ($\Phi$-C$_3$H$_5$Z) | | 32.8 |
| Isomers of C$_{10}$H$_{10}$ (C$_{10}$H$_{10}$#) or | | 31.7 | 3-phenyl-1-propen-3-yl radical or 3-phenyl-2-propen-1-yl radical $\Phi$-C$_3$H$_4$Y Resonance stabilized | | 58.5 |
| Indane | | 22.5 | | | |
| Indene | | 34.9 | Indenyl radical (indenyl) Resonance stabilized | | 67.2 |
| Naphthalene | | 31.4 | Naphthenyl radicals (naphthenyl) | | 90.7 |

**FIGURE CAPTIONS**

Figure 1: Scheme of the apparatus of low-pressure laminar premixed flame. The thick lines correspond to the heated lines.

Figure 2: Typical chromatogram of light compounds obtained at a distance of 1.40 mm from the burner at 1380 K (oven temperature program: 333 K during 10 min, then a rise of 5 K/min until 523 K).

Figure 3: Typical chromatogram of heavy compounds obtained at a distance of 1.31 mm from the burner at 1310 K (oven temperature program: 313 K during 30 min, then a rise of 5 K/min until 453 K).

Figure 4: Temperature profiles: experimental measurements performed without and with the sampling probe and profile used for simulation.

Figure 5: Profiles of the mole fractions of both hydrocarbon reactants including a computed comparison between doped and undoped flames. Points are experiments and lines simulations. In the case of methane, the full lines correspond to the flame seeded with n-butylbenzene and the broken one to a simulated flame of pure methane at $\Phi=0.74$ (see text).

Figure 6: Profiles of the mole fractions of oxygen and the main oxygenated products including a computed comparison between doped and undoped flames. Points are experiments and lines simulations. Full lines correspond to the flame seeded with n-butylbenzene and broken lines to a simulated flame of pure methane at $\Phi=0.74$ (see text).

Figure 7: Profiles of the mole fractions of hydrogen and $C_2$ species including a computed comparison between doped and undoped flames. Points are experiments and lines simulations. Full lines correspond to the flame seeded with n-butylbenzene and the broken lines to a simulated flame of pure methane at $\Phi=0.74$ (see text).

Figure 8: Profiles of the mole fractions of $C_3$ species including a computed comparison between doped and undoped flames. Points are experiments and lines simulations. Full lines correspond to the flame seeded with n-butylbenzene and the broken lines to a simulated flame of pure methane at $\Phi$=0.74 (see text).

Figure 9: Profiles of the mole fractions of $C_4$ species. Points are experiments and lines simulations.

Figure 10: Profiles of the mole fractions of $C_5$-$C_6$ non-aromatic species. Points are experiments and lines simulations.

Figure 11: Profiles of the mole fractions of oxygenated $C_1$-$C_3$ species. Points are experiments and lines simulations. Full lines correspond to the flame seeded with butybenzene and the broken lines to a simulated flame of pure methane at $\Phi$=0.74 (see text).

Figure 12: Profiles of the mole fractions of $C_6$-$C_8$ aromatic species. Points are experiments and lines simulations.

Figure 13: Profiles of the mole fractions of $C_9$-$C_{10}$ monoaromatic species. Points are experiments and lines simulations.

Figure 14: Profiles of the mole fractions of biaromatic species. Points are experiments and lines simulations.

Figure 15: Profiles of the mole fractions of oxygenated aromatic species. Points are experiments and lines simulations.

Figure 16: Computed species mole fractions versus residence time during the oxidation of n-butylbenzene in a flow reactor at 1060 K, P= 1atm and $\Phi$ = 0.98, with an initial concentration of hydrocarbon of 620 ppm [6]. Points are experiments and lines simulations.

Figure 17: Flow rate analysis for the consumption of the n-butylbenzene for a distance of 1.15 mm from the burner corresponding to a temperature of 1144 K and a 87 % conversion of butylbenzene. The size of the arrows is proportional to the relative flow rates.

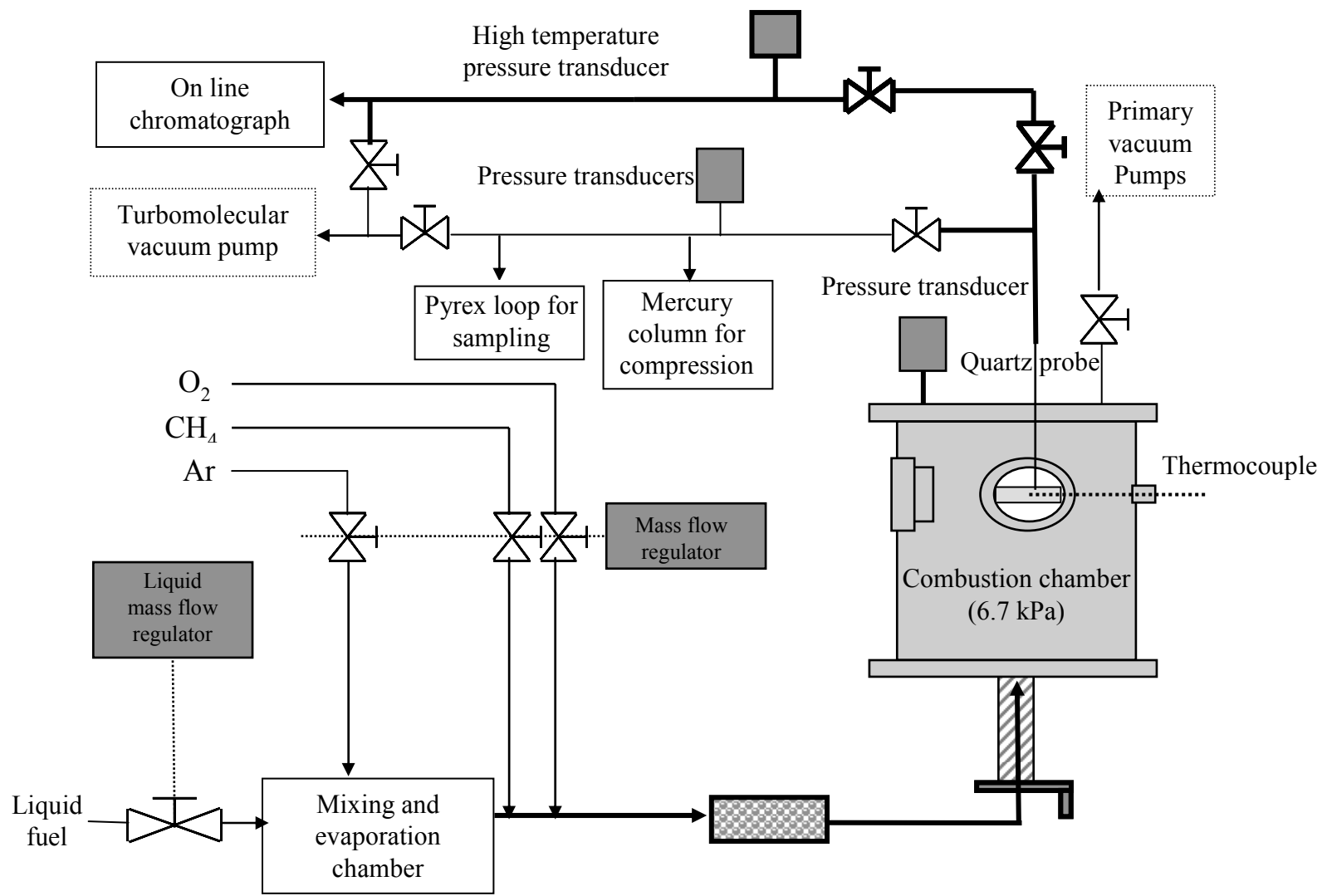

High temperature
pressure transducer

On line
chromatograph

Primary
vacuum
Pumps

Pressure transducers

Turbomolecular
vacuum pump

Pyrex loop for
sampling

Mercury
column for
compression

Pressure transducer

Quartz probe

O₂

CH₄

Ar

Mass flow
regulator

Thermocouple

Liquid
mass flow
regulator

Combustion chamber
(6.7 kPa)

Liquid
fuel

Mixing and
evaporation
chamber

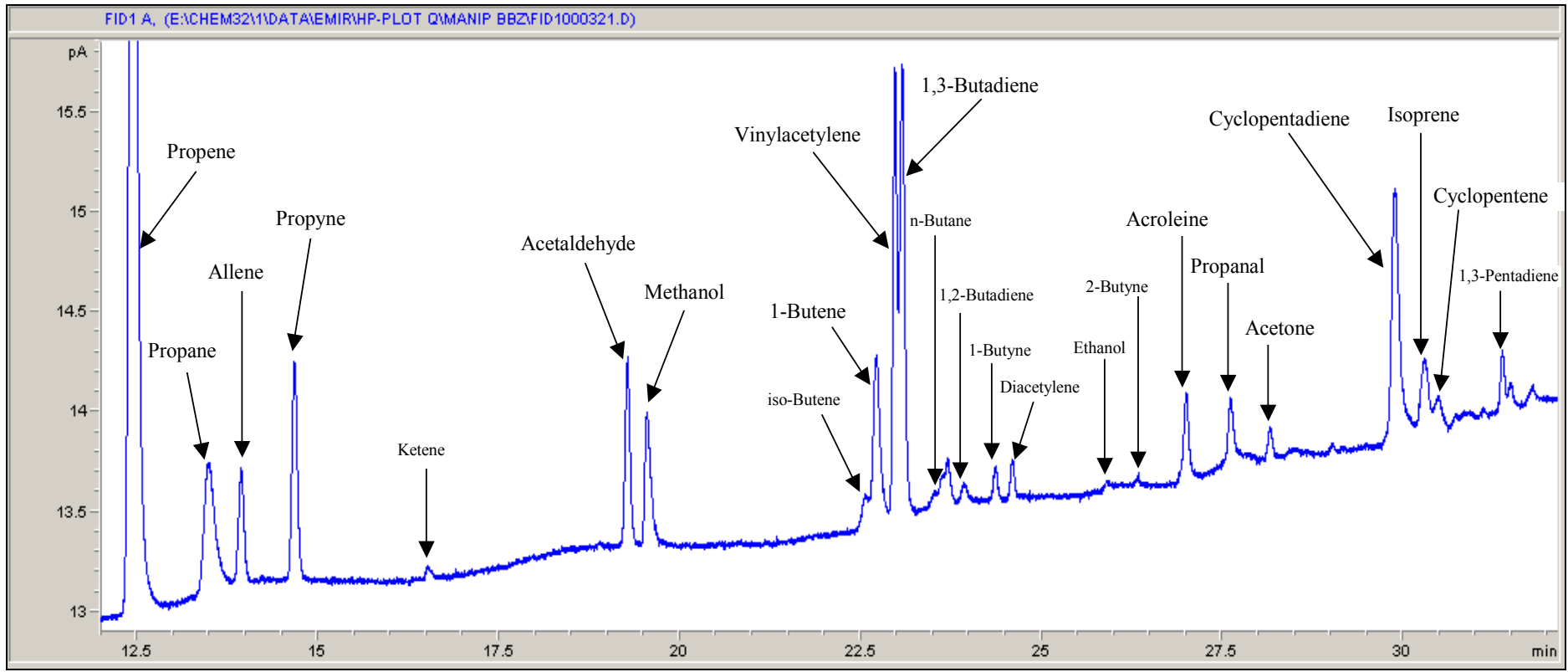

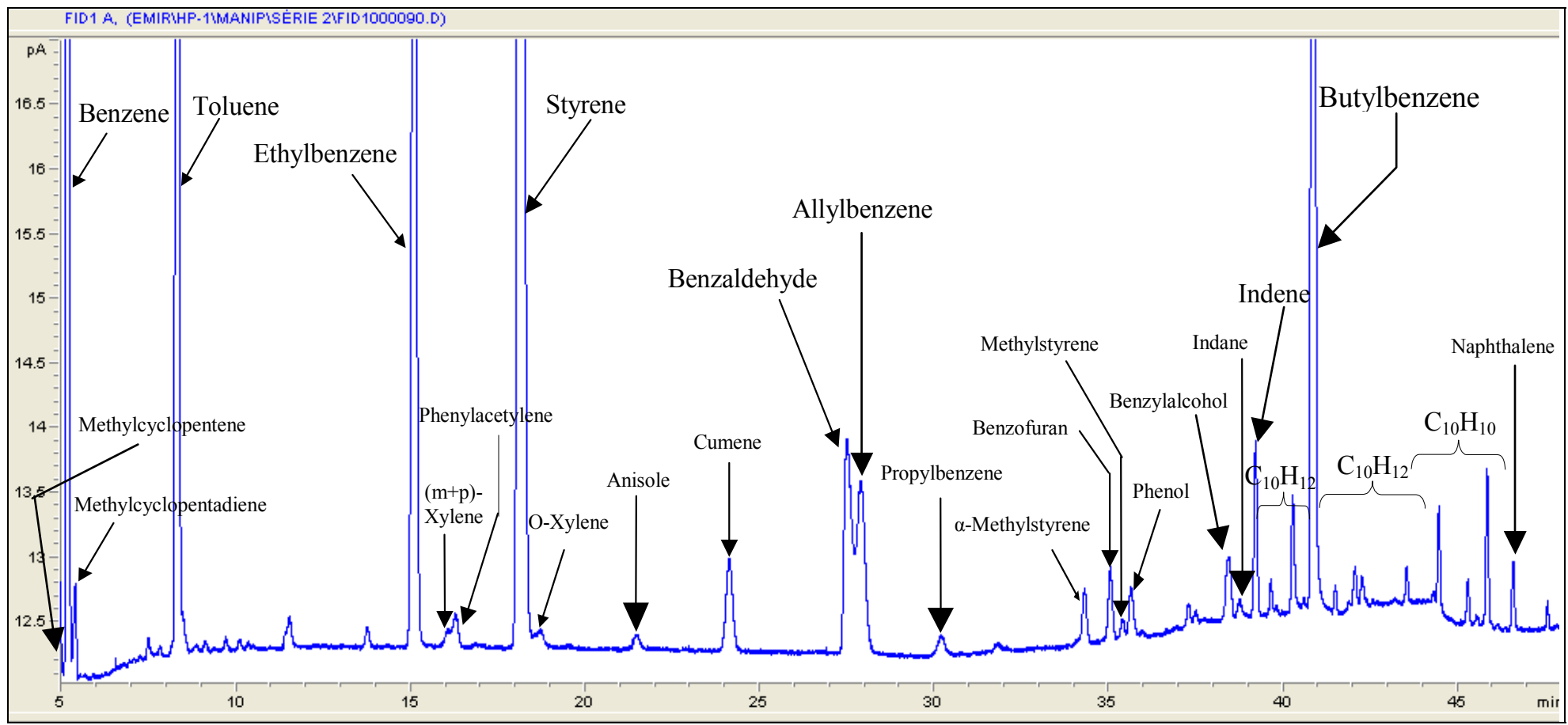

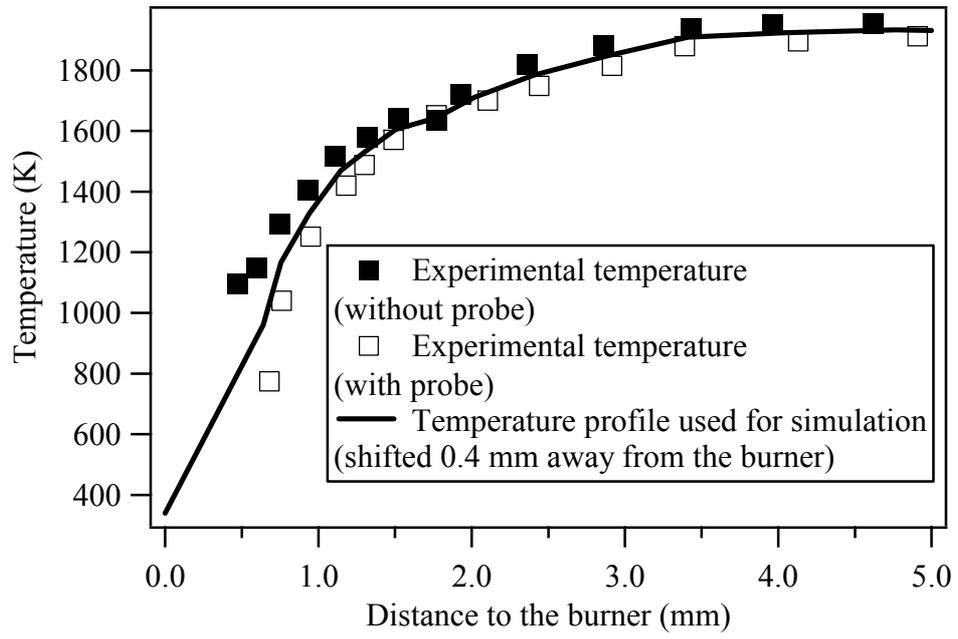

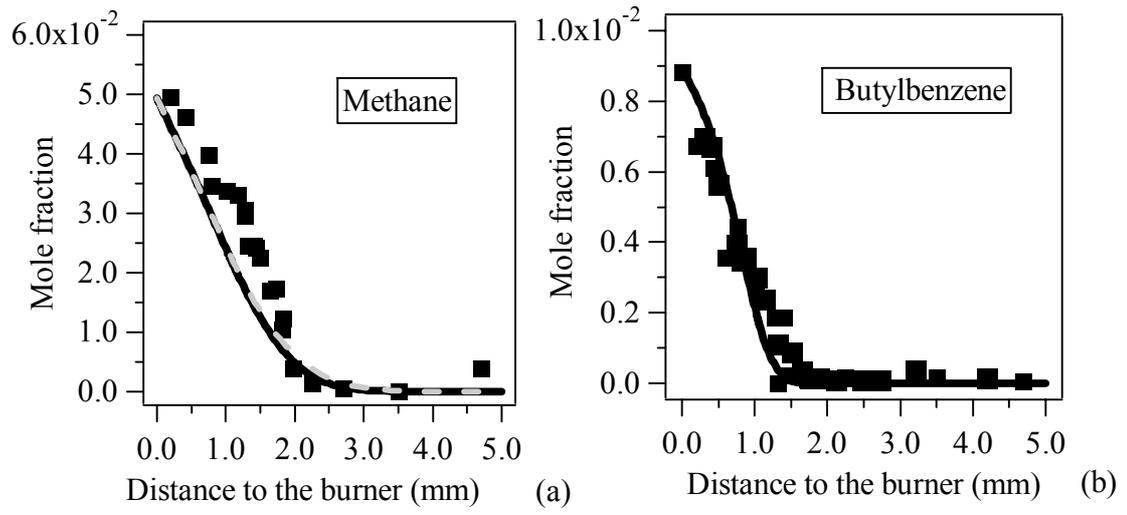

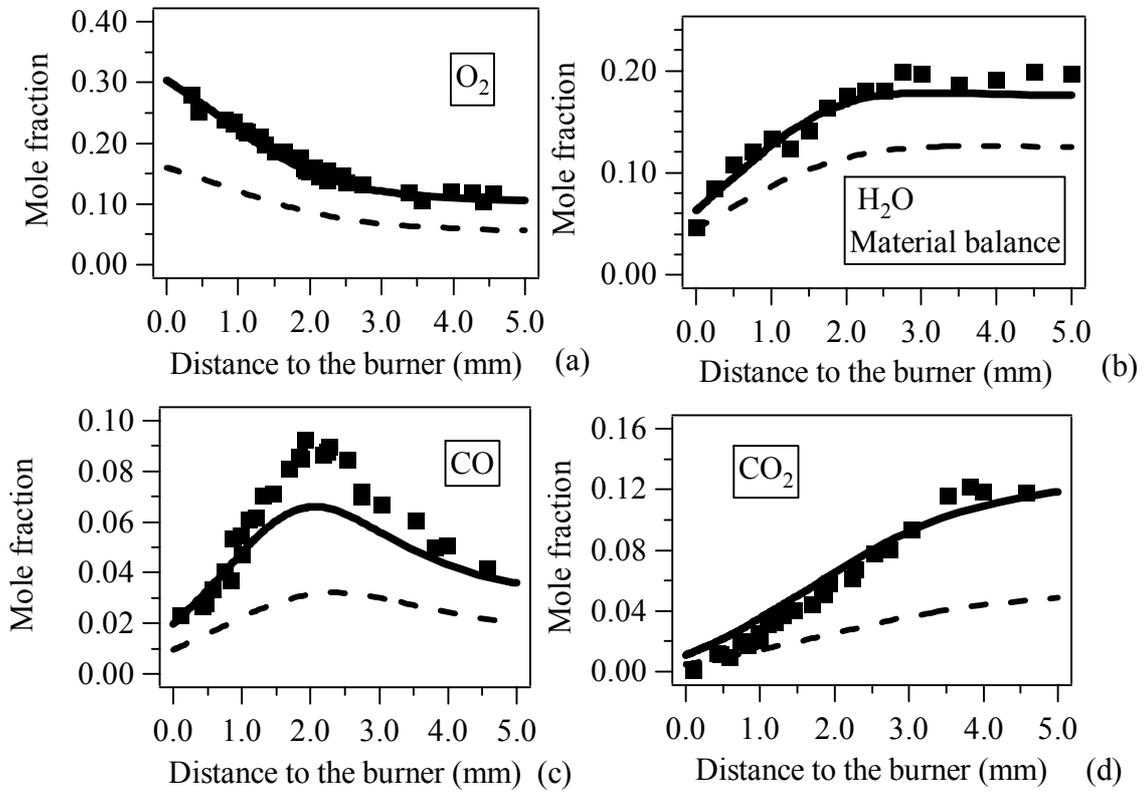

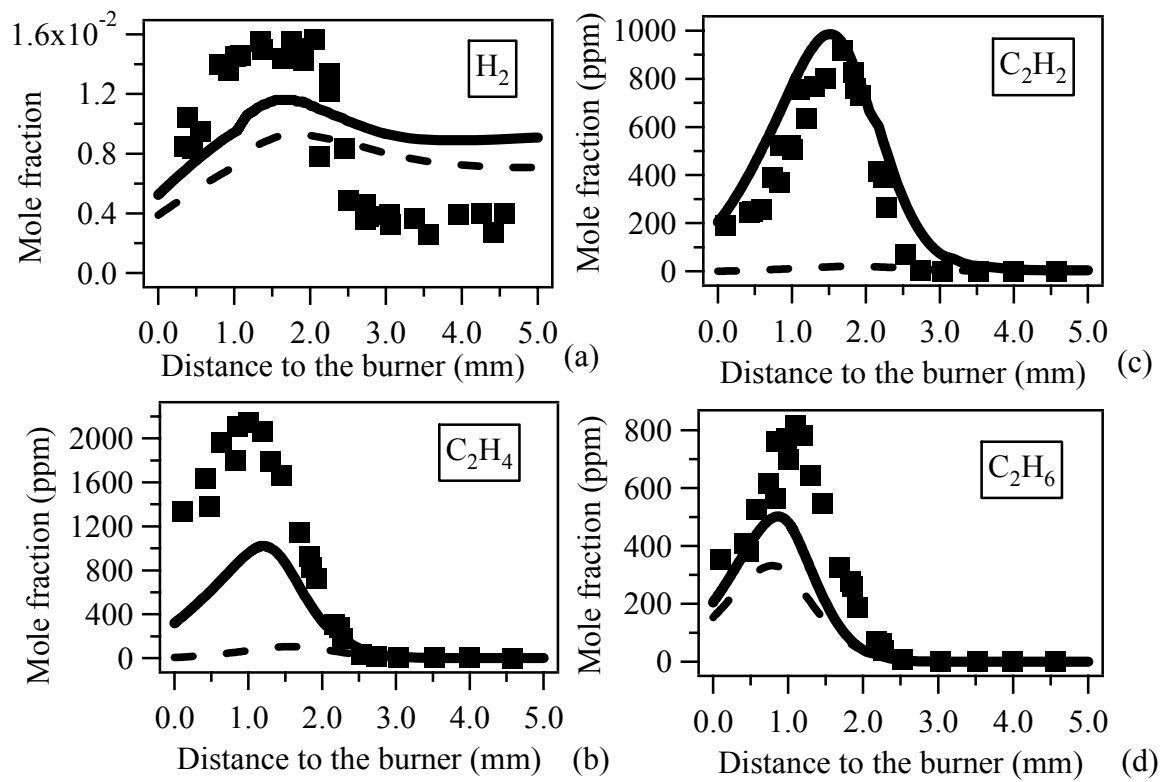

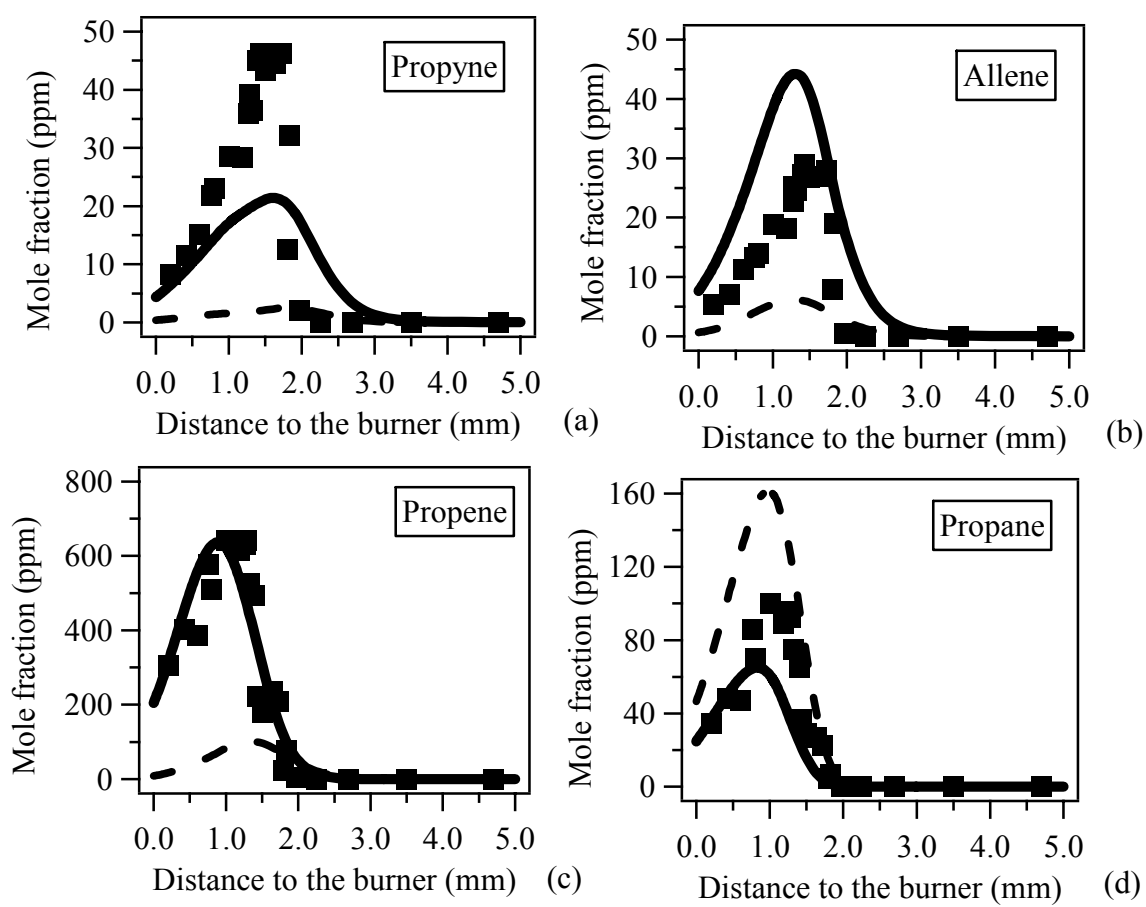

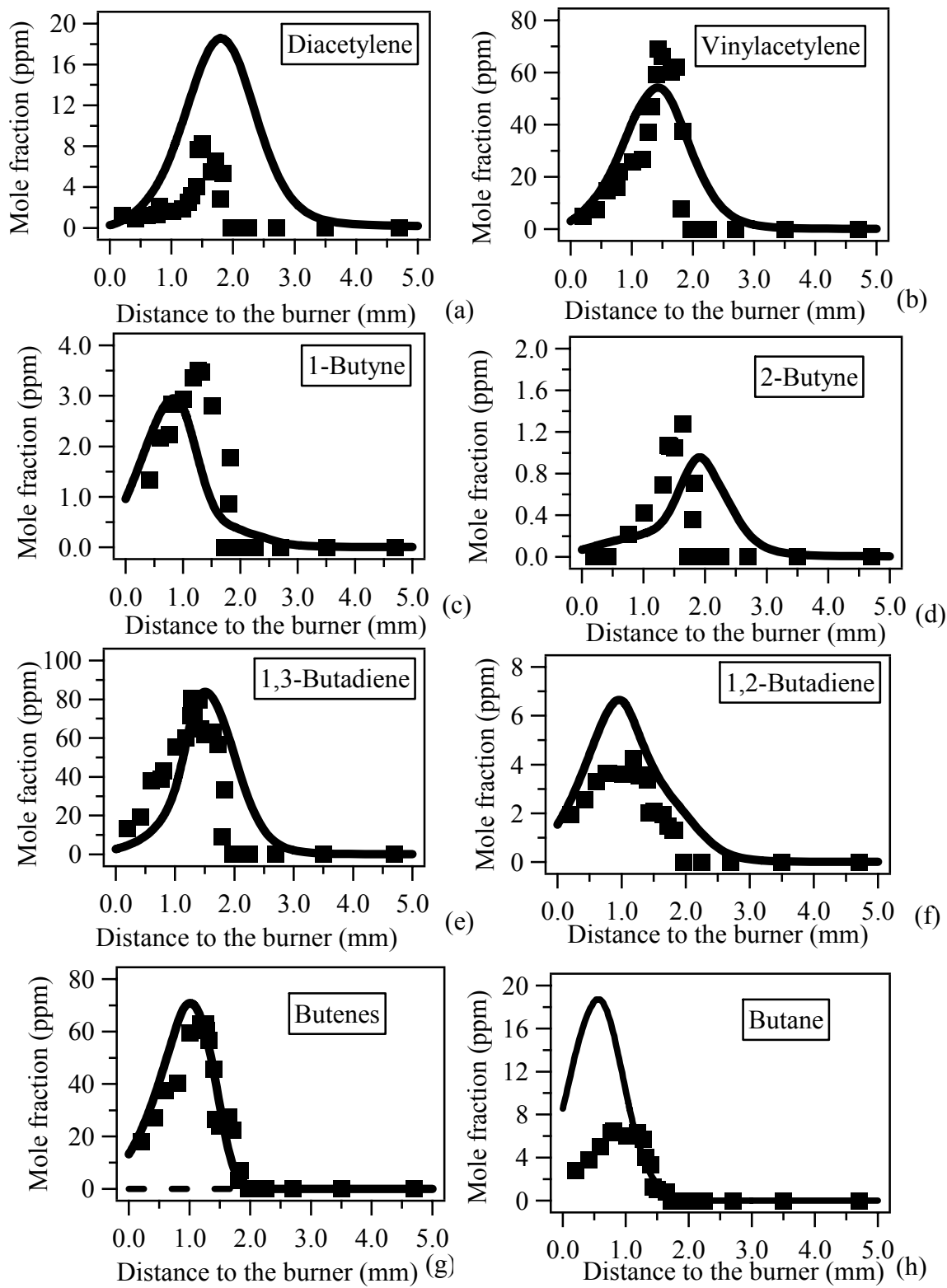

Figure 10

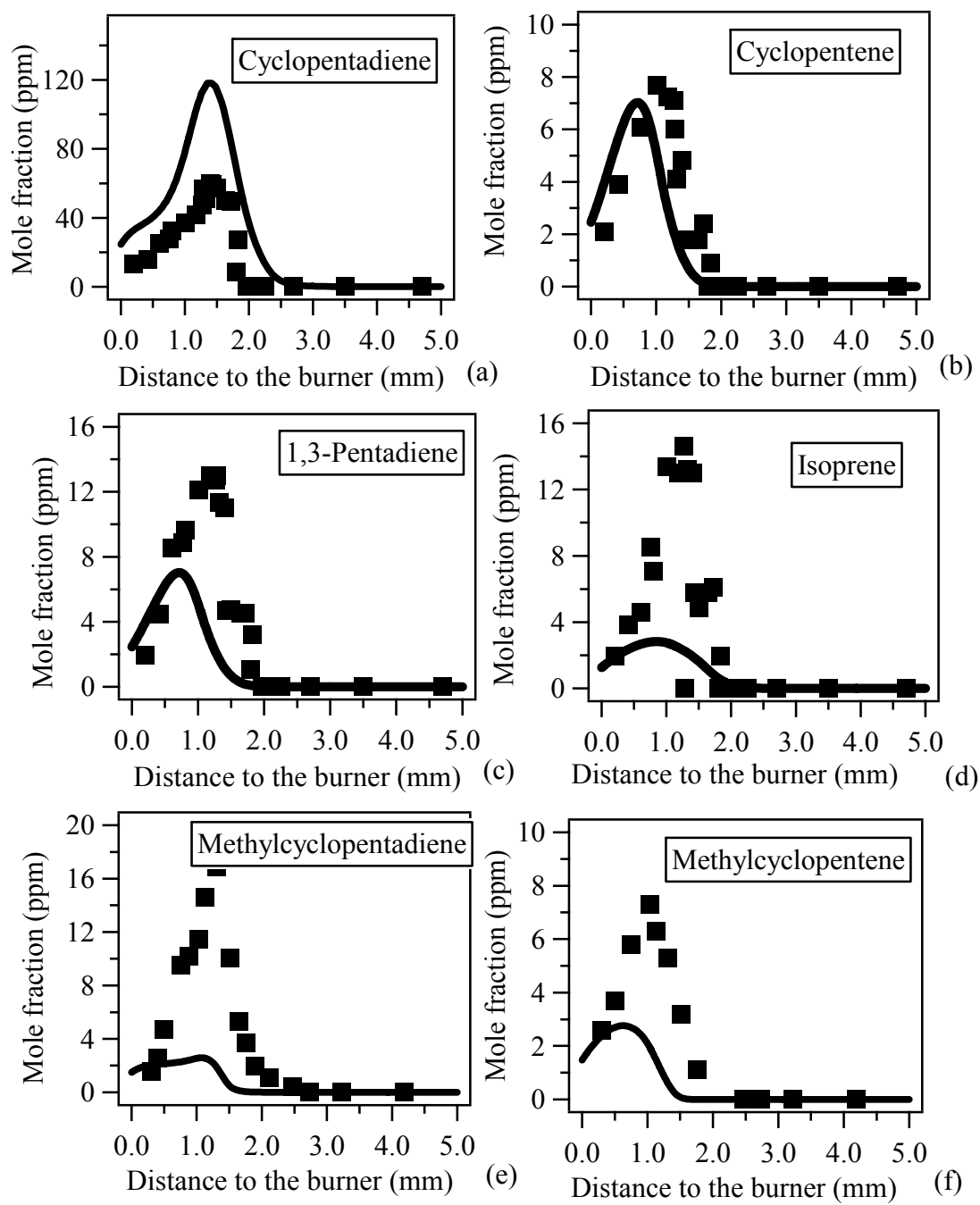

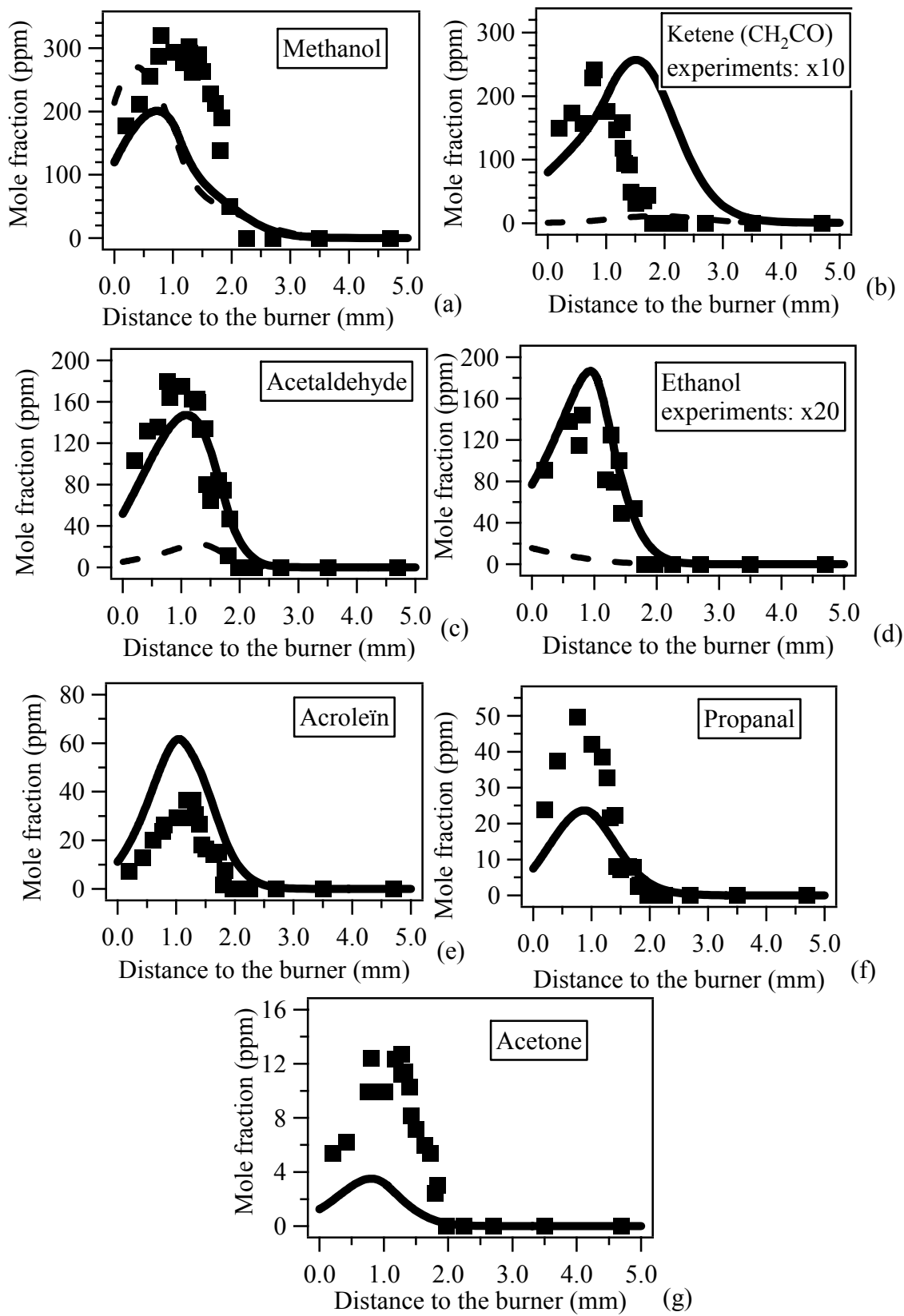

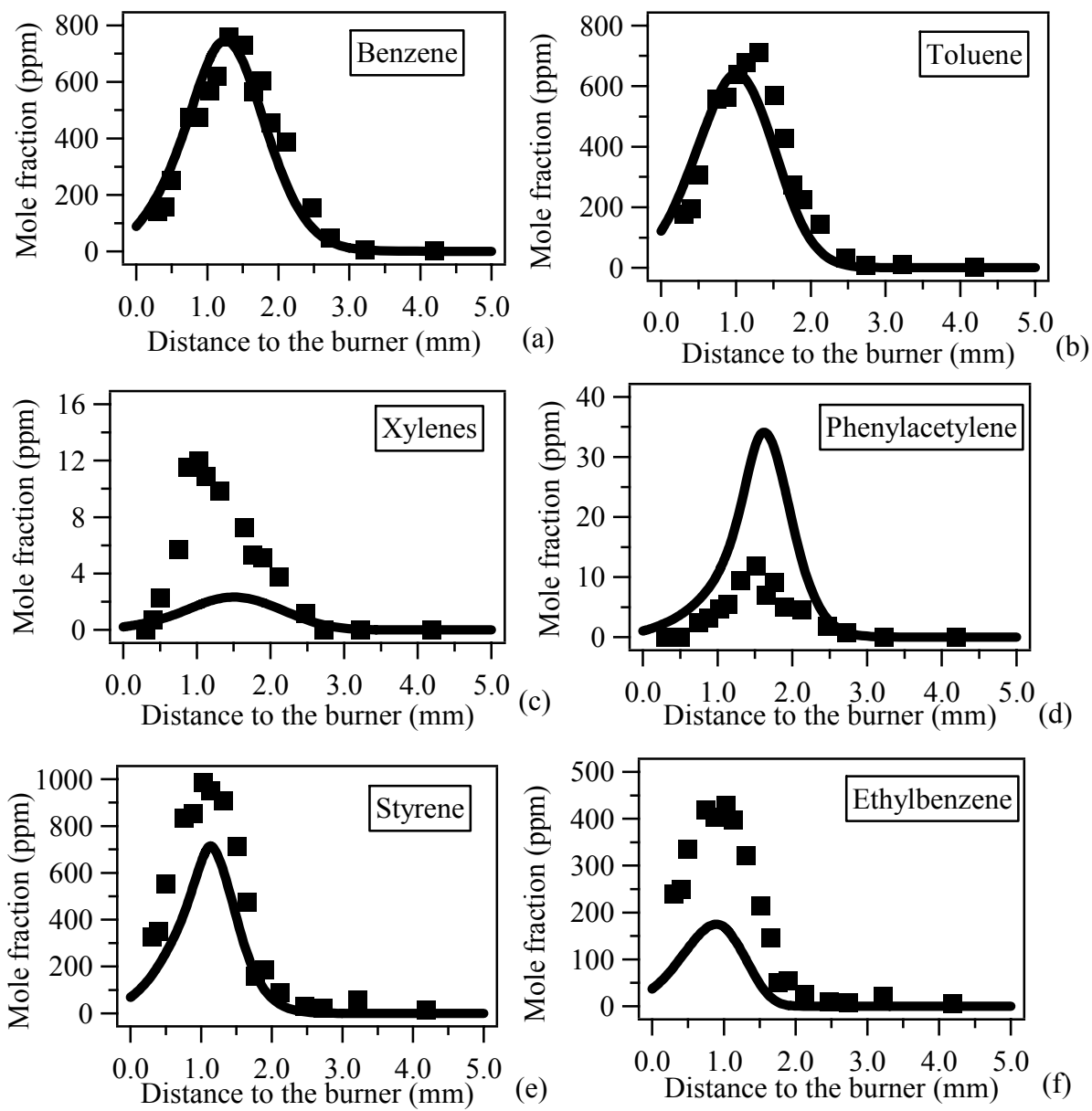

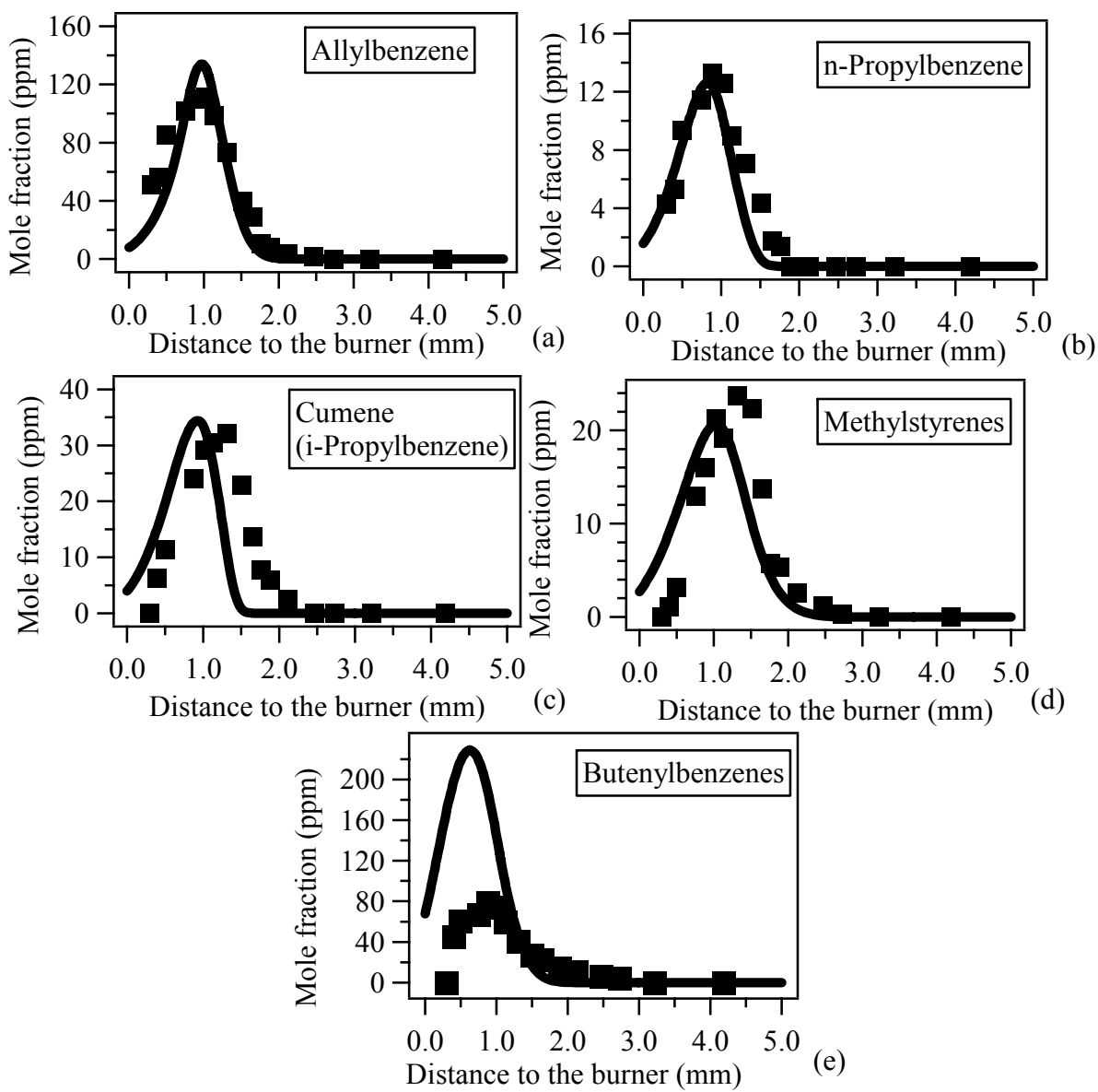

(a)

(b)

(c)

(d)

(e)

Figure 14

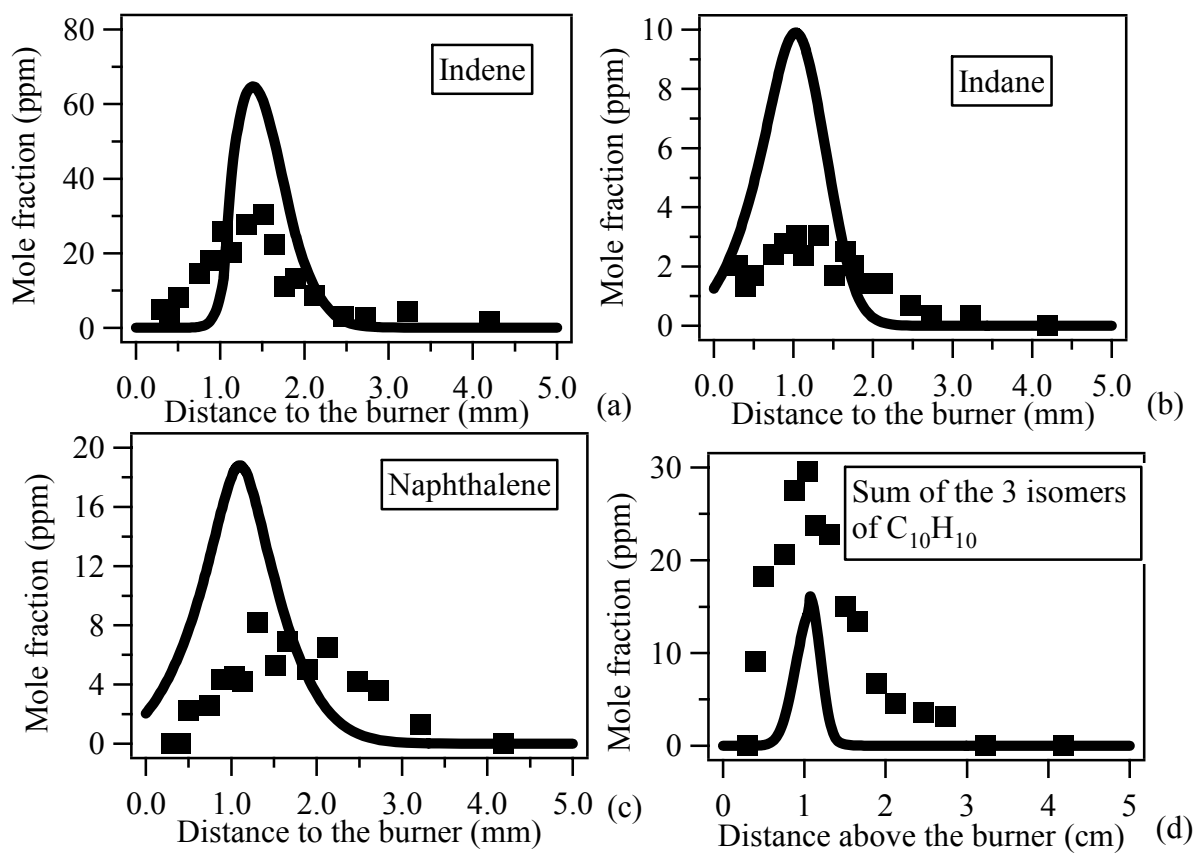

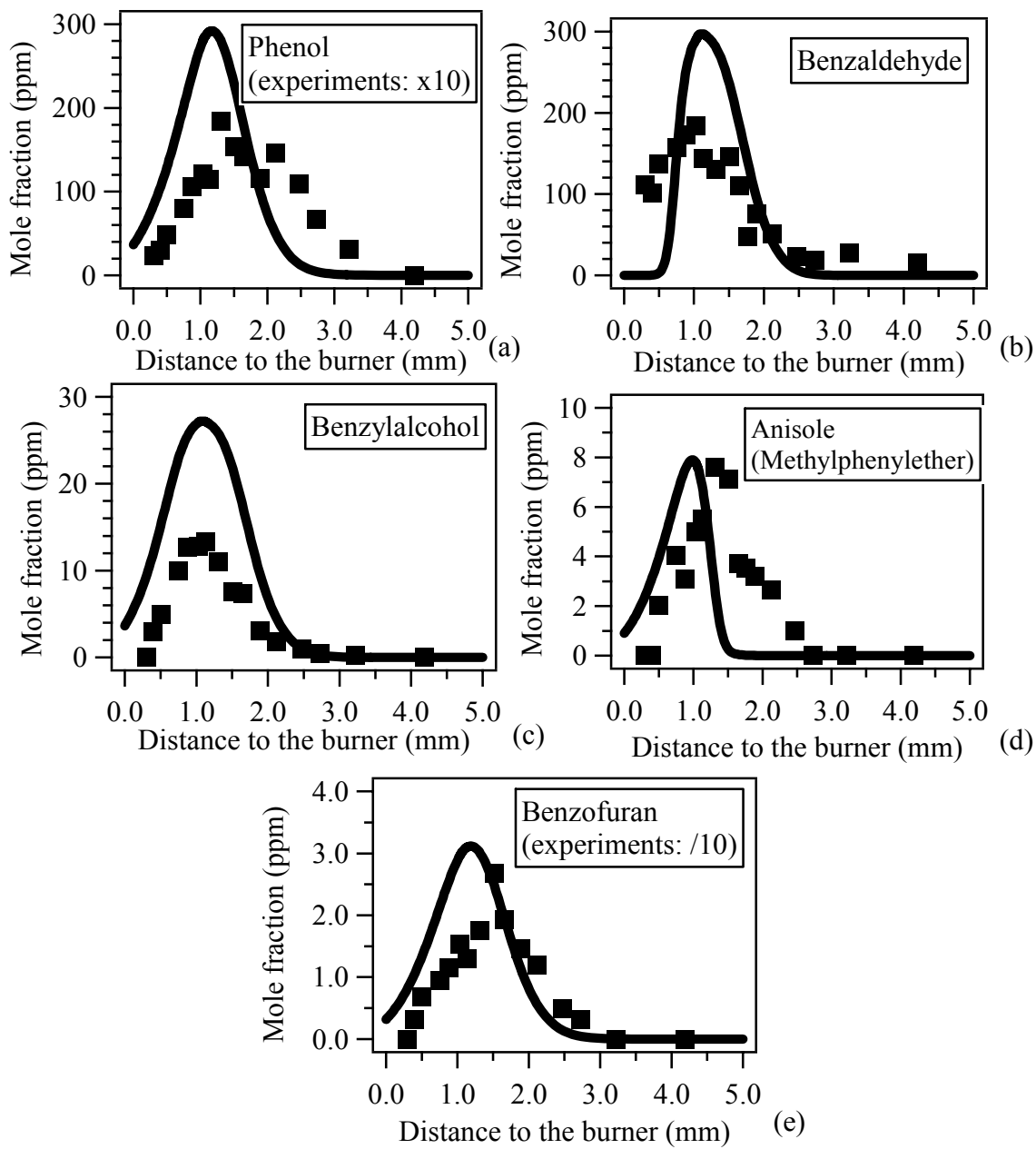

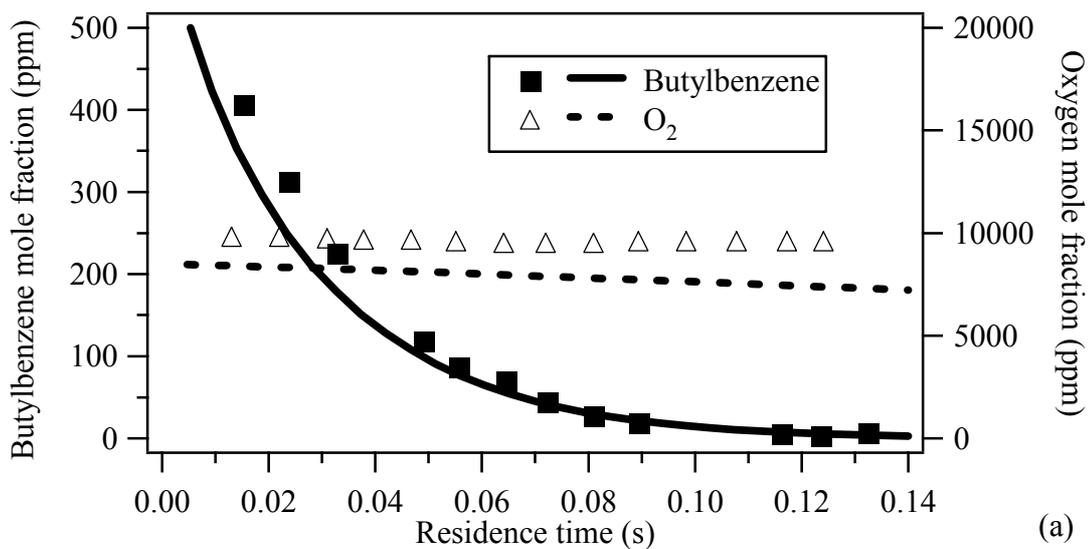

(a)

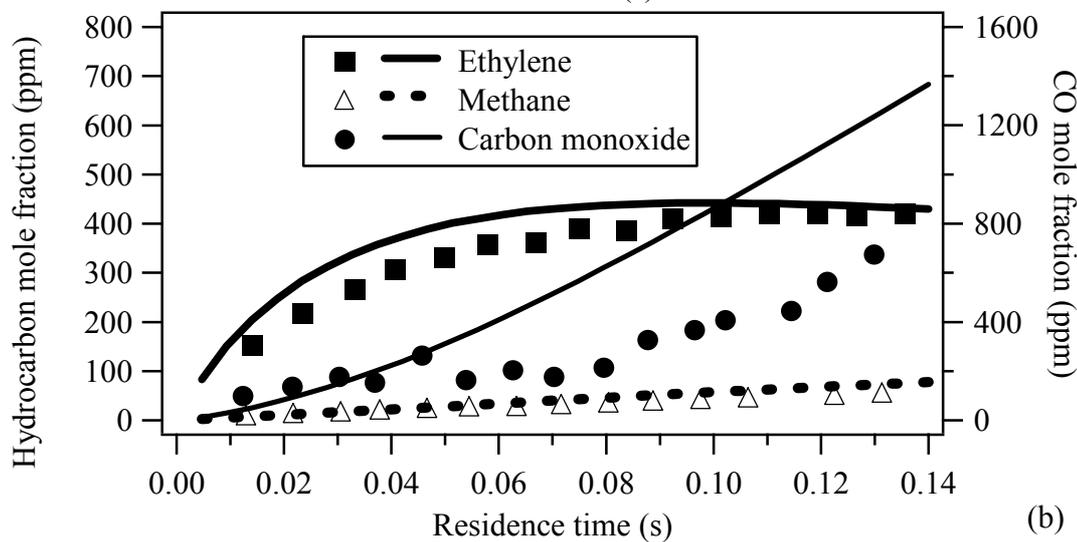

(b)

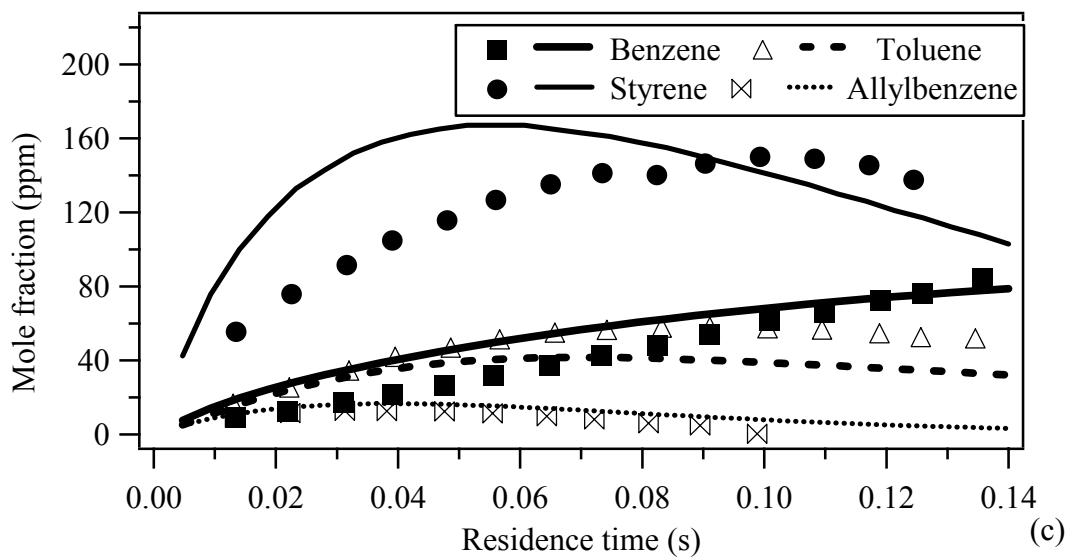

(c)